
\catcode`\@=11


\message{Loading jyTeX fonts...}



\font\vptrm=cmr5 \font\vptmit=cmmi5 \font\vptsy=cmsy5 \font\vptbf=cmbx5

\skewchar\vptmit='177 \skewchar\vptsy='60 \fontdimen16
\vptsy=\the\fontdimen17 \vptsy

\def\vpt{\ifmmode\err@badsizechange\else
     \@mathfontinit
     \textfont0=\vptrm  \scriptfont0=\vptrm  \scriptscriptfont0=\vptrm
     \textfont1=\vptmit \scriptfont1=\vptmit \scriptscriptfont1=\vptmit
     \textfont2=\vptsy  \scriptfont2=\vptsy  \scriptscriptfont2=\vptsy
     \textfont3=\xptex  \scriptfont3=\xptex  \scriptscriptfont3=\xptex
     \textfont\bffam=\vptbf
     \scriptfont\bffam=\vptbf
     \scriptscriptfont\bffam=\vptbf
     \@fontstyleinit
     \def\rm{\vptrm\fam=\z@}%
     \def\bf{\vptbf\fam=\bffam}%
     \def\oldstyle{\vptmit\fam=\@ne}%
     \rm\fi}


\font\viptrm=cmr6 \font\viptmit=cmmi6 \font\viptsy=cmsy6
\font\viptbf=cmbx6

\skewchar\viptmit='177 \skewchar\viptsy='60 \fontdimen16
\viptsy=\the\fontdimen17 \viptsy

\def\vipt{\ifmmode\err@badsizechange\else
     \@mathfontinit
     \textfont0=\viptrm  \scriptfont0=\vptrm  \scriptscriptfont0=\vptrm
     \textfont1=\viptmit \scriptfont1=\vptmit \scriptscriptfont1=\vptmit
     \textfont2=\viptsy  \scriptfont2=\vptsy  \scriptscriptfont2=\vptsy
     \textfont3=\xptex   \scriptfont3=\xptex  \scriptscriptfont3=\xptex
     \textfont\bffam=\viptbf
     \scriptfont\bffam=\vptbf
     \scriptscriptfont\bffam=\vptbf
     \@fontstyleinit
     \def\rm{\viptrm\fam=\z@}%
     \def\bf{\viptbf\fam=\bffam}%
     \def\oldstyle{\viptmit\fam=\@ne}%
     \rm\fi}

\font\viiptrm=cmr7 \font\viiptmit=cmmi7 \font\viiptsy=cmsy7
\font\viiptit=cmti7 \font\viiptbf=cmbx7

\skewchar\viiptmit='177 \skewchar\viiptsy='60 \fontdimen16
\viiptsy=\the\fontdimen17 \viiptsy

\def\viipt{\ifmmode\err@badsizechange\else
     \@mathfontinit
     \textfont0=\viiptrm  \scriptfont0=\vptrm  \scriptscriptfont0=\vptrm
     \textfont1=\viiptmit \scriptfont1=\vptmit \scriptscriptfont1=\vptmit
     \textfont2=\viiptsy  \scriptfont2=\vptsy  \scriptscriptfont2=\vptsy
     \textfont3=\xptex    \scriptfont3=\xptex  \scriptscriptfont3=\xptex
     \textfont\itfam=\viiptit
     \scriptfont\itfam=\viiptit
     \scriptscriptfont\itfam=\viiptit
     \textfont\bffam=\viiptbf
     \scriptfont\bffam=\vptbf
     \scriptscriptfont\bffam=\vptbf
     \@fontstyleinit
     \def\rm{\viiptrm\fam=\z@}%
     \def\it{\viiptit\fam=\itfam}%
     \def\bf{\viiptbf\fam=\bffam}%
     \def\oldstyle{\viiptmit\fam=\@ne}%
     \rm\fi}


\font\viiiptrm=cmr8 \font\viiiptmit=cmmi8 \font\viiiptsy=cmsy8
\font\viiiptit=cmti8
\font\viiiptbf=cmbx8

\skewchar\viiiptmit='177 \skewchar\viiiptsy='60 \fontdimen16
\viiiptsy=\the\fontdimen17 \viiiptsy

\def\viiipt{\ifmmode\err@badsizechange\else
     \@mathfontinit
     \textfont0=\viiiptrm  \scriptfont0=\viptrm  \scriptscriptfont0=\vptrm
     \textfont1=\viiiptmit \scriptfont1=\viptmit \scriptscriptfont1=\vptmit
     \textfont2=\viiiptsy  \scriptfont2=\viptsy  \scriptscriptfont2=\vptsy
     \textfont3=\xptex     \scriptfont3=\xptex   \scriptscriptfont3=\xptex
     \textfont\itfam=\viiiptit
     \scriptfont\itfam=\viiptit
     \scriptscriptfont\itfam=\viiptit
     \textfont\bffam=\viiiptbf
     \scriptfont\bffam=\viptbf
     \scriptscriptfont\bffam=\vptbf
     \@fontstyleinit
     \def\rm{\viiiptrm\fam=\z@}%
     \def\it{\viiiptit\fam=\itfam}%
     \def\bf{\viiiptbf\fam=\bffam}%
     \def\oldstyle{\viiiptmit\fam=\@ne}%
     \rm\fi}


\def\getixpt{%
     \font\ixptrm=cmr9
     \font\ixptmit=cmmi9
     \font\ixptsy=cmsy9
     \font\ixptit=cmti9
     \font\ixptbf=cmbx9
     \skewchar\ixptmit='177 \skewchar\ixptsy='60
     \fontdimen16 \ixptsy=\the\fontdimen17 \ixptsy}

\def\ixpt{\ifmmode\err@badsizechange\else
     \@mathfontinit
     \textfont0=\ixptrm  \scriptfont0=\viiptrm  \scriptscriptfont0=\vptrm
     \textfont1=\ixptmit \scriptfont1=\viiptmit \scriptscriptfont1=\vptmit
     \textfont2=\ixptsy  \scriptfont2=\viiptsy  \scriptscriptfont2=\vptsy
     \textfont3=\xptex   \scriptfont3=\xptex    \scriptscriptfont3=\xptex
     \textfont\itfam=\ixptit
     \scriptfont\itfam=\viiptit
     \scriptscriptfont\itfam=\viiptit
     \textfont\bffam=\ixptbf
     \scriptfont\bffam=\viiptbf
     \scriptscriptfont\bffam=\vptbf
     \@fontstyleinit
     \def\rm{\ixptrm\fam=\z@}%
     \def\it{\ixptit\fam=\itfam}%
     \def\bf{\ixptbf\fam=\bffam}%
     \def\oldstyle{\ixptmit\fam=\@ne}%
     \rm\fi}


\font\xptrm=cmr10 \font\xptmit=cmmi10 \font\xptsy=cmsy10
\font\xptex=cmex10 \font\xptit=cmti10 \font\xptsl=cmsl10
\font\xptbf=cmbx10 \font\xpttt=cmtt10 \font\xptss=cmss10
\font\xptsc=cmcsc10 \font\xptbfs=cmb10 \font\xptbmit=cmmib10

\skewchar\xptmit='177 \skewchar\xptbmit='177 \skewchar\xptsy='60
\fontdimen16 \xptsy=\the\fontdimen17 \xptsy

\def\xpt{\ifmmode\err@badsizechange\else
     \@mathfontinit
     \textfont0=\xptrm  \scriptfont0=\viiptrm  \scriptscriptfont0=\vptrm
     \textfont1=\xptmit \scriptfont1=\viiptmit \scriptscriptfont1=\vptmit
     \textfont2=\xptsy  \scriptfont2=\viiptsy  \scriptscriptfont2=\vptsy
     \textfont3=\xptex  \scriptfont3=\xptex    \scriptscriptfont3=\xptex
     \textfont\itfam=\xptit
     \scriptfont\itfam=\viiptit
     \scriptscriptfont\itfam=\viiptit
     \textfont\bffam=\xptbf
     \scriptfont\bffam=\viiptbf
     \scriptscriptfont\bffam=\vptbf
     \textfont\bfsfam=\xptbfs
     \scriptfont\bfsfam=\viiptbf
     \scriptscriptfont\bfsfam=\vptbf
     \textfont\bmitfam=\xptbmit
     \scriptfont\bmitfam=\viiptmit
     \scriptscriptfont\bmitfam=\vptmit
     \@fontstyleinit
     \def\rm{\xptrm\fam=\z@}%
     \def\it{\xptit\fam=\itfam}%
     \def\sl{\xptsl}%
     \def\bf{\xptbf\fam=\bffam}%
     \def\tt{\xpttt}%
     \def\ss{\xptss}%
     \def\sc{\xptsc}%
     \def\bfs{\xptbfs\fam=\bfsfam}%
     \def\bmit{\fam=\bmitfam}%
     \def\oldstyle{\xptmit\fam=\@ne}%
     \rm\fi}


\def\getxipt{%
     \font\xiptrm=cmr10  scaled\magstephalf
     \font\xiptmit=cmmi10 scaled\magstephalf
     \font\xiptsy=cmsy10 scaled\magstephalf
     \font\xiptex=cmex10 scaled\magstephalf
     \font\xiptit=cmti10 scaled\magstephalf
     \font\xiptsl=cmsl10 scaled\magstephalf
     \font\xiptbf=cmbx10 scaled\magstephalf
     \font\xipttt=cmtt10 scaled\magstephalf
     \font\xiptss=cmss10 scaled\magstephalf
     \skewchar\xiptmit='177 \skewchar\xiptsy='60
     \fontdimen16 \xiptsy=\the\fontdimen17 \xiptsy}

\def\xipt{\ifmmode\err@badsizechange\else
     \@mathfontinit
     \textfont0=\xiptrm  \scriptfont0=\viiiptrm  \scriptscriptfont0=\viptrm
     \textfont1=\xiptmit \scriptfont1=\viiiptmit \scriptscriptfont1=\viptmit
     \textfont2=\xiptsy  \scriptfont2=\viiiptsy  \scriptscriptfont2=\viptsy
     \textfont3=\xiptex  \scriptfont3=\xptex     \scriptscriptfont3=\xptex
     \textfont\itfam=\xiptit
     \scriptfont\itfam=\viiiptit
     \scriptscriptfont\itfam=\viiptit
     \textfont\bffam=\xiptbf
     \scriptfont\bffam=\viiiptbf
     \scriptscriptfont\bffam=\viptbf
     \@fontstyleinit
     \def\rm{\xiptrm\fam=\z@}%
     \def\it{\xiptit\fam=\itfam}%
     \def\sl{\xiptsl}%
     \def\bf{\xiptbf\fam=\bffam}%
     \def\tt{\xipttt}%
     \def\ss{\xiptss}%
     \def\oldstyle{\xiptmit\fam=\@ne}%
     \rm\fi}


\font\xiiptrm=cmr12 \font\xiiptmit=cmmi12 \font\xiiptsy=cmsy10
scaled\magstep1 \font\xiiptex=cmex10  scaled\magstep1
\font\xiiptit=cmti12 \font\xiiptsl=cmsl12 \font\xiiptbf=cmbx12
\font\xiiptss=cmss12 \font\xiiptsc=cmcsc10 scaled\magstep1
\font\xiiptbfs=cmb10  scaled\magstep1 \font\xiiptbmit=cmmib10
scaled\magstep1

\skewchar\xiiptmit='177 \skewchar\xiiptbmit='177 \skewchar\xiiptsy='60
\fontdimen16 \xiiptsy=\the\fontdimen17 \xiiptsy

\def\xiipt{\ifmmode\err@badsizechange\else
     \@mathfontinit
     \textfont0=\xiiptrm  \scriptfont0=\viiiptrm  \scriptscriptfont0=\viptrm
     \textfont1=\xiiptmit \scriptfont1=\viiiptmit \scriptscriptfont1=\viptmit
     \textfont2=\xiiptsy  \scriptfont2=\viiiptsy  \scriptscriptfont2=\viptsy
     \textfont3=\xiiptex  \scriptfont3=\xptex     \scriptscriptfont3=\xptex
     \textfont\itfam=\xiiptit
     \scriptfont\itfam=\viiiptit
     \scriptscriptfont\itfam=\viiptit
     \textfont\bffam=\xiiptbf
     \scriptfont\bffam=\viiiptbf
     \scriptscriptfont\bffam=\viptbf
     \textfont\bfsfam=\xiiptbfs
     \scriptfont\bfsfam=\viiiptbf
     \scriptscriptfont\bfsfam=\viptbf
     \textfont\bmitfam=\xiiptbmit
     \scriptfont\bmitfam=\viiiptmit
     \scriptscriptfont\bmitfam=\viptmit
     \@fontstyleinit
     \def\rm{\xiiptrm\fam=\z@}%
     \def\it{\xiiptit\fam=\itfam}%
     \def\sl{\xiiptsl}%
     \def\bf{\xiiptbf\fam=\bffam}%
     \def\tt{\xiipttt}%
     \def\ss{\xiiptss}%
     \def\sc{\xiiptsc}%
     \def\bfs{\xiiptbfs\fam=\bfsfam}%
     \def\bmit{\fam=\bmitfam}%
     \def\oldstyle{\xiiptmit\fam=\@ne}%
     \rm\fi}


\def\getxiiipt{%
     \font\xiiiptrm=cmr12  scaled\magstephalf
     \font\xiiiptmit=cmmi12 scaled\magstephalf
     \font\xiiiptsy=cmsy9  scaled\magstep2
     \font\xiiiptit=cmti12 scaled\magstephalf
     \font\xiiiptsl=cmsl12 scaled\magstephalf
     \font\xiiiptbf=cmbx12 scaled\magstephalf
     \font\xiiipttt=cmtt12 scaled\magstephalf
     \font\xiiiptss=cmss12 scaled\magstephalf
     \skewchar\xiiiptmit='177 \skewchar\xiiiptsy='60
     \fontdimen16 \xiiiptsy=\the\fontdimen17 \xiiiptsy}

\def\xiiipt{\ifmmode\err@badsizechange\else
     \@mathfontinit
     \textfont0=\xiiiptrm  \scriptfont0=\xptrm  \scriptscriptfont0=\viiptrm
     \textfont1=\xiiiptmit \scriptfont1=\xptmit \scriptscriptfont1=\viiptmit
     \textfont2=\xiiiptsy  \scriptfont2=\xptsy  \scriptscriptfont2=\viiptsy
     \textfont3=\xivptex   \scriptfont3=\xptex  \scriptscriptfont3=\xptex
     \textfont\itfam=\xiiiptit
     \scriptfont\itfam=\xptit
     \scriptscriptfont\itfam=\viiptit
     \textfont\bffam=\xiiiptbf
     \scriptfont\bffam=\xptbf
     \scriptscriptfont\bffam=\viiptbf
     \@fontstyleinit
     \def\rm{\xiiiptrm\fam=\z@}%
     \def\it{\xiiiptit\fam=\itfam}%
     \def\sl{\xiiiptsl}%
     \def\bf{\xiiiptbf\fam=\bffam}%
     \def\tt{\xiiipttt}%
     \def\ss{\xiiiptss}%
     \def\oldstyle{\xiiiptmit\fam=\@ne}%
     \rm\fi}


\font\xivptrm=cmr12   scaled\magstep1 \font\xivptmit=cmmi12
scaled\magstep1 \font\xivptsy=cmsy10  scaled\magstep2
\font\xivptex=cmex10  scaled\magstep2 \font\xivptit=cmti12
scaled\magstep1 \font\xivptsl=cmsl12  scaled\magstep1
\font\xivptbf=cmbx12  scaled\magstep1
\font\xivptss=cmss12  scaled\magstep1 \font\xivptsc=cmcsc10
scaled\magstep2 \font\xivptbfs=cmb10  scaled\magstep2
\font\xivptbmit=cmmib10 scaled\magstep2

\skewchar\xivptmit='177 \skewchar\xivptbmit='177 \skewchar\xivptsy='60
\fontdimen16 \xivptsy=\the\fontdimen17 \xivptsy

\def\xivpt{\ifmmode\err@badsizechange\else
     \@mathfontinit
     \textfont0=\xivptrm  \scriptfont0=\xptrm  \scriptscriptfont0=\viiptrm
     \textfont1=\xivptmit \scriptfont1=\xptmit \scriptscriptfont1=\viiptmit
     \textfont2=\xivptsy  \scriptfont2=\xptsy  \scriptscriptfont2=\viiptsy
     \textfont3=\xivptex  \scriptfont3=\xptex  \scriptscriptfont3=\xptex
     \textfont\itfam=\xivptit
     \scriptfont\itfam=\xptit
     \scriptscriptfont\itfam=\viiptit
     \textfont\bffam=\xivptbf
     \scriptfont\bffam=\xptbf
     \scriptscriptfont\bffam=\viiptbf
     \textfont\bfsfam=\xivptbfs
     \scriptfont\bfsfam=\xptbfs
     \scriptscriptfont\bfsfam=\viiptbf
     \textfont\bmitfam=\xivptbmit
     \scriptfont\bmitfam=\xptbmit
     \scriptscriptfont\bmitfam=\viiptmit
     \@fontstyleinit
     \def\rm{\xivptrm\fam=\z@}%
     \def\it{\xivptit\fam=\itfam}%
     \def\sl{\xivptsl}%
     \def\bf{\xivptbf\fam=\bffam}%
     \def\tt{\xivpttt}%
     \def\ss{\xivptss}%
     \def\sc{\xivptsc}%
     \def\bfs{\xivptbfs\fam=\bfsfam}%
     \def\bmit{\fam=\bmitfam}%
     \def\oldstyle{\xivptmit\fam=\@ne}%
     \rm\fi}


\font\xviiptrm=cmr17 \font\xviiptmit=cmmi12 scaled\magstep2
\font\xviiptsy=cmsy10 scaled\magstep3 \font\xviiptex=cmex10
scaled\magstep3 \font\xviiptit=cmti12 scaled\magstep2
\font\xviiptbf=cmbx12 scaled\magstep2 \font\xviiptbfs=cmb10
scaled\magstep3

\skewchar\xviiptmit='177 \skewchar\xviiptsy='60 \fontdimen16
\xviiptsy=\the\fontdimen17 \xviiptsy

\def\xviipt{\ifmmode\err@badsizechange\else
     \@mathfontinit
     \textfont0=\xviiptrm  \scriptfont0=\xiiptrm  \scriptscriptfont0=\viiiptrm
     \textfont1=\xviiptmit \scriptfont1=\xiiptmit \scriptscriptfont1=\viiiptmit
     \textfont2=\xviiptsy  \scriptfont2=\xiiptsy  \scriptscriptfont2=\viiiptsy
     \textfont3=\xviiptex  \scriptfont3=\xiiptex  \scriptscriptfont3=\xptex
     \textfont\itfam=\xviiptit
     \scriptfont\itfam=\xiiptit
     \scriptscriptfont\itfam=\viiiptit
     \textfont\bffam=\xviiptbf
     \scriptfont\bffam=\xiiptbf
     \scriptscriptfont\bffam=\viiiptbf
     \textfont\bfsfam=\xviiptbfs
     \scriptfont\bfsfam=\xiiptbfs
     \scriptscriptfont\bfsfam=\viiiptbf
     \@fontstyleinit
     \def\rm{\xviiptrm\fam=\z@}%
     \def\it{\xviiptit\fam=\itfam}%
     \def\bf{\xviiptbf\fam=\bffam}%
     \def\bfs{\xviiptbfs\fam=\bfsfam}%
     \def\oldstyle{\xviiptmit\fam=\@ne}%
     \rm\fi}


\font\xxiptrm=cmr17  scaled\magstep1


\def\xxipt{\ifmmode\err@badsizechange\else
     \@mathfontinit
     \@fontstyleinit
     \def\rm{\xxiptrm\fam=\z@}%
     \rm\fi}


\font\xxvptrm=cmr17  scaled\magstep2


\def\xxvpt{\ifmmode\err@badsizechange\else
     \@mathfontinit
     \@fontstyleinit
     \def\rm{\xxvptrm\fam=\z@}%
     \rm\fi}




\message{Loading jyTeX macros...}

\message{modifications to plain.tex,}


\def\newcount{\alloc@0\count\countdef\insc@unt}
\def\newdimen{\alloc@1\dimen\dimendef\insc@unt}
\def\newskip{\alloc@2\skip\skipdef\insc@unt}
\def\newmuskip{\alloc@3\muskip\muskipdef\@cclvi}
\def\newbox{\alloc@4\box\chardef\insc@unt}
\def\newtoks{\alloc@5\toks\toksdef\@cclvi}
\def\newhelp#1#2{\newtoks#1\global#1\expandafter{\csname#2\endcsname}}
\def\newread{\alloc@6\read\chardef\sixt@@n}
\def\newwrite{\alloc@7\write\chardef\sixt@@n}
\def\newfam{\alloc@8\fam\chardef\sixt@@n}
\def\newinsert#1{\global\advance\insc@unt by\m@ne
     \ch@ck0\insc@unt\count
     \ch@ck1\insc@unt\dimen
     \ch@ck2\insc@unt\skip
     \ch@ck4\insc@unt\box
     \allocationnumber=\insc@unt
     \global\chardef#1=\allocationnumber
     \wlog{\string#1=\string\insert\the\allocationnumber}}
\def\newif#1{\count@\escapechar \escapechar\m@ne
     \expandafter\expandafter\expandafter
          \xdef\@if#1{true}{\let\noexpand#1=\noexpand\iftrue}%
     \expandafter\expandafter\expandafter
          \xdef\@if#1{false}{\let\noexpand#1=\noexpand\iffalse}%
     \global\@if#1{false}\escapechar=\count@}


\newlinechar=`\^^J
\overfullrule=0pt




\let\itfam=\undefined

\let\bffam=\undefined

\count18=3


\chardef\sharps="19


\mathchardef\alpha="710B \mathchardef\beta="710C \mathchardef\gamma="710D
\mathchardef\delta="710E \mathchardef\epsilon="710F
\mathchardef\zeta="7110 \mathchardef\eta="7111 \mathchardef\theta="7112
\mathchardef\iota="7113 \mathchardef\kappa="7114
\mathchardef\lambda="7115 \mathchardef\mu="7116 \mathchardef\nu="7117
\mathchardef\xi="7118 \mathchardef\pi="7119 \mathchardef\rho="711A
\mathchardef\sigma="711B \mathchardef\tau="711C
\mathchardef\upsilon="711D \mathchardef\phi="711E \mathchardef\chi="711F
\mathchardef\psi="7120 \mathchardef\omega="7121
\mathchardef\varepsilon="7122 \mathchardef\vartheta="7123
\mathchardef\varpi="7124 \mathchardef\varrho="7125
\mathchardef\varsigma="7126 \mathchardef\varphi="7127
\mathchardef\imath="717B \mathchardef\jmath="717C \mathchardef\ell="7160
\mathchardef\wp="717D \mathchardef\partial="7140 \mathchardef\flat="715B
\mathchardef\natural="715C \mathchardef\sharp="715D



\def\angle{{\vbox{\ialign{$\m@th\scriptstyle##$\crcr
     \not\mathrel{\mkern14mu}\crcr
     \noalign{\nointerlineskip}
     \mkern2.5mu\leaders\hrule height.34\rp@\hfill\mkern2.5mu\crcr}}}}
\def\vdots{\vbox{\baselineskip4\rp@ \lineskiplimit\z@
     \kern6\rp@\hbox{.}\hbox{.}\hbox{.}}}
\def\ddots{\mathinner{\mkern1mu\raise7\rp@\vbox{\kern7\rp@\hbox{.}}\mkern2mu
     \raise4\rp@\hbox{.}\mkern2mu\raise\rp@\hbox{.}\mkern1mu}}
\def\overrightarrow#1{\vbox{\ialign{##\crcr
     \rightarrowfill\crcr
     \noalign{\kern-\rp@\nointerlineskip}
     $\hfil\displaystyle{#1}\hfil$\crcr}}}
\def\overleftarrow#1{\vbox{\ialign{##\crcr
     \leftarrowfill\crcr
     \noalign{\kern-\rp@\nointerlineskip}
     $\hfil\displaystyle{#1}\hfil$\crcr}}}
\def\overbrace#1{\mathop{\vbox{\ialign{##\crcr
     \noalign{\kern3\rp@}
     \downbracefill\crcr
     \noalign{\kern3\rp@\nointerlineskip}
     $\hfil\displaystyle{#1}\hfil$\crcr}}}\limits}
\def\underbrace#1{\mathop{\vtop{\ialign{##\crcr
     $\hfil\displaystyle{#1}\hfil$\crcr
     \noalign{\kern3\rp@\nointerlineskip}
     \upbracefill\crcr
     \noalign{\kern3\rp@}}}}\limits}
\def\big#1{{\hbox{$\left#1\vbox to8.5\rp@ {}\right.\n@space$}}}
\def\Big#1{{\hbox{$\left#1\vbox to11.5\rp@ {}\right.\n@space$}}}
\def\bigg#1{{\hbox{$\left#1\vbox to14.5\rp@ {}\right.\n@space$}}}
\def\Bigg#1{{\hbox{$\left#1\vbox to17.5\rp@ {}\right.\n@space$}}}
\def\@vereq#1#2{\lower.5\rp@\vbox{\baselineskip\z@skip\lineskip-.5\rp@
     \ialign{$\m@th#1\hfil##\hfil$\crcr#2\crcr=\crcr}}}
\def\rlh@#1{\vcenter{\hbox{\ooalign{\raise2\rp@
     \hbox{$#1\rightharpoonup$}\crcr
     $#1\leftharpoondown$}}}}
\def\bordermatrix#1{\begingroup\m@th
     \setbox\z@\vbox{%
          \def\cr{\crcr\noalign{\kern2\rp@\global\let\cr\endline}}%
          \ialign{$##$\hfil\kern2\rp@\kern\p@renwd
               &\thinspace\hfil$##$\hfil&&\quad\hfil$##$\hfil\crcr
               \omit\strut\hfil\crcr
               \noalign{\kern-\baselineskip}%
               #1\crcr\omit\strut\cr}}%
     \setbox\tw@\vbox{\unvcopy\z@\global\setbox\@ne\lastbox}%
     \setbox\tw@\hbox{\unhbox\@ne\unskip\global\setbox\@ne\lastbox}%
     \setbox\tw@\hbox{$\kern\wd\@ne\kern-\p@renwd\left(\kern-\wd\@ne
          \global\setbox\@ne\vbox{\box\@ne\kern2\rp@}%
          \vcenter{\kern-\ht\@ne\unvbox\z@\kern-\baselineskip}%
          \,\right)$}%
     \null\;\vbox{\kern\ht\@ne\box\tw@}\endgroup}
\def\endinsert{\egroup
     \if@mid\dimen@\ht\z@
          \advance\dimen@\dp\z@
          \advance\dimen@12\rp@
          \advance\dimen@\pagetotal
          \ifdim\dimen@>\pagegoal\@midfalse\p@gefalse\fi
     \fi
     \if@mid\bigskip\box\z@
          \bigbreak
     \else\insert\topins{\penalty100 \splittopskip\z@skip
               \splitmaxdepth\maxdimen\floatingpenalty\z@
               \ifp@ge\dimen@\dp\z@
                    \vbox to\vsize{\unvbox\z@\kern-\dimen@}%
               \else\box\z@\nobreak\bigskip
               \fi}%
     \fi
     \endgroup}


\def\cases#1{\left\{\,\vcenter{\m@th
     \ialign{$##\hfil$&\quad##\hfil\crcr#1\crcr}}\right.}
\def\matrix#1{\null\,\vcenter{\m@th
     \ialign{\hfil$##$\hfil&&\quad\hfil$##$\hfil\crcr
          \mathstrut\crcr
          \noalign{\kern-\baselineskip}
          #1\crcr
          \mathstrut\crcr
          \noalign{\kern-\baselineskip}}}\,}


\newif\ifraggedbottom

\def\raggedbottom{\ifraggedbottom\else
     \advance\topskip by\z@ plus60pt \raggedbottomtrue\fi}%
\def\normalbottom{\ifraggedbottom
     \advance\topskip by\z@ plus-60pt \raggedbottomfalse\fi}

\message{hacks,}


\toksdef\toks@i=1 \toksdef\toks@ii=2


\def\TeX{T\kern-.1667em \lower.5ex \hbox{E}\kern-.125em X\null}
\def\jyTeX{{\leavevmode
     \raise.587ex \hbox{\it\j}\kern-.1em \lower.048ex \hbox{\it y}\kern-.12em
     \TeX}}

\let\then=\iftrue
\def\ifnoarg#1\then{\def\hack@{#1}\ifx\hack@\empty}
\def\ifundefined#1\then{%
     \expandafter\ifx\csname\expandafter\blank\string#1\endcsname\relax}
\def\useif#1\then{\csname#1\endcsname}
\def\usename#1{\csname#1\endcsname}
\def\useafter#1#2{\expandafter#1\csname#2\endcsname}

\long\def\loop#1\repeat{\def\@iterate{#1\expandafter\@iterate\fi}\@iterate
     \let\@iterate=\relax}

\let\TeXend=\end
\def\begin#1{\begingroup\def\@@blockname{#1}\usename{begin#1}}
\def\end#1{\usename{end#1}\def\hack@{#1}%
     \ifx\@@blockname\hack@
          \endgroup
     \else\err@badgroup\hack@\@@blockname
     \fi}
\def\@@blockname{}

\def\defaultoption[#1]#2{%
     \def\hack@{\ifx\hack@ii[\toks@={#2}\else\toks@={#2[#1]}\fi\the\toks@}%
     \futurelet\hack@ii\hack@}

\def\markup#1{\let\@@marksf=\empty
     \ifhmode\edef\@@marksf{\spacefactor=\the\spacefactor\relax}\/\fi
     ${}^{\hbox{\subscriptfonts#1}}$\@@marksf}


\newtoks\shortyear
\newtoks\militaryhour
\newtoks\standardhour
\newtoks\minute
\newtoks\amorpm

\def\settime{\count@=\time\divide\count@ by60
     \militaryhour=\expandafter{\number\count@}%
     {\multiply\count@ by-60 \advance\count@ by\time
          \xdef\hack@{\ifnum\count@<10 0\fi\number\count@}}%
     \minute=\expandafter{\hack@}%
     \ifnum\count@<12
          \amorpm={am}
     \else\amorpm={pm}
          \ifnum\count@>12 \advance\count@ by-12 \fi
     \fi
     \standardhour=\expandafter{\number\count@}%
     \def\hack@19##1##2{\shortyear={##1##2}}%
          \expandafter\hack@\the\year}

\def\monthword#1{%
     \ifcase#1
          $\bullet$\err@badcountervalue{monthword}%
          \or January\or February\or March\or April\or May\or June%
          \or July\or August\or September\or October\or November\or December%
     \else$\bullet$\err@badcountervalue{monthword}%
     \fi}

\def\monthabbr#1{%
     \ifcase#1
          $\bullet$\err@badcountervalue{monthabbr}%
          \or Jan\or Feb\or Mar\or Apr\or May\or Jun%
          \or Jul\or Aug\or Sep\or Oct\or Nov\or Dec%
     \else$\bullet$\err@badcountervalue{monthabbr}%
     \fi}

\def\militarytime{\the\militaryhour:\the\minute}
\def\standardtime{\the\standardhour:\the\minute}


\def\@setnumstyle#1#2{\expandafter\global\expandafter\expandafter
     \expandafter\let\expandafter\expandafter
     \csname @\expandafter\blank\string#1style\endcsname
     \csname#2\endcsname}
\def\numstyle#1{\usename{@\expandafter\blank\string#1style}#1}
\def\ifblank#1\then{\useafter\ifx{@\expandafter\blank\string#1}\blank}

\def\blank#1{}

\def\Roman#1{\expandafter\uppercase\expandafter{\romannumeral#1}}
\def\alphabetic#1{%
     \ifcase#1
          $\bullet$\err@badcountervalue{alphabetic}%
          \or a\or b\or c\or d\or e\or f\or g\or h\or i\or j\or k\or l\or m%
          \or n\or o\or p\or q\or r\or s\or t\or u\or v\or w\or x\or y\or z%
     \else$\bullet$\err@badcountervalue{alphabetic}%
     \fi}
\def\Alphabetic#1{\expandafter\uppercase\expandafter{\alphabetic{#1}}}
\def\symbols#1{%
     \ifcase#1
          $\bullet$\err@badcountervalue{symbols}%
          \or*\or\dag\or\ddag\or\S\or$\|$%
          \or**\or\dag\dag\or\ddag\ddag\or\S\S\or$\|\|$%
     \else$\bullet$\err@badcountervalue{symbols}%
     \fi}


\catcode`\^^?=13 \def^^?{\relax}

\def\trimleading#1\to#2{\edef#2{#1}%
     \expandafter\@trimleading\expandafter#2#2^^?^^?}
\def\@trimleading#1#2#3^^?{\ifx#2^^?\def#1{}\else\def#1{#2#3}\fi}

\def\trimtrailing#1\to#2{\edef#2{#1}%
     \expandafter\@trimtrailing\expandafter#2#2^^? ^^?\relax}
\def\@trimtrailing#1#2 ^^?#3{\ifx#3\relax\toks@={}%
     \else\def#1{#2}\toks@={\trimtrailing#1\to#1}\fi
     \the\toks@}

\def\trim#1\to#2{\trimleading#1\to#2\trimtrailing#2\to#2}

\catcode`\^^?=15


\long\def\additemL#1\to#2{\toks@={\^^\{#1}}\toks@ii=\expandafter{#2}%
     \xdef#2{\the\toks@\the\toks@ii}}

\long\def\additemR#1\to#2{\toks@={\^^\{#1}}\toks@ii=\expandafter{#2}%
     \xdef#2{\the\toks@ii\the\toks@}}

\def\getitemL#1\to#2{\expandafter\@getitemL#1\hack@#1#2}
\def\@getitemL\^^\#1#2\hack@#3#4{\def#4{#1}\def#3{#2}}

\message{font macros,}


\newdimen\rp@
\newcount\@@sizeindex \@@sizeindex=0
\newcount\@@factori
\newcount\@@factorii
\newcount\@@factoriii
\newcount\@@factoriv

\countdef\maxfam=18
\newfam\itfam
\newfam\bffam
\newfam\bfsfam
\newfam\bmitfam

\def\@mathfontinit{\count@=4
     \loop\textfont\count@=\nullfont
          \scriptfont\count@=\nullfont
          \scriptscriptfont\count@=\nullfont
          \ifnum\count@<\maxfam\advance\count@ by\@ne
     \repeat}

\def\@fontstyleinit{%
     \def\it{\err@fontnotavailable\it}%
     \def\bf{\err@fontnotavailable\bf}%
     \def\bfs{\err@bfstobf}%
     \def\bmit{\err@fontnotavailable\bmit}%
     \def\sc{\err@fontnotavailable\sc}%
     \def\sl{\err@sltoit}%
     \def\ss{\err@fontnotavailable\ss}%
     \def\tt{\err@fontnotavailable\tt}}

\def\@parameterinit#1{\rm\rp@=.1em \@getscaling{#1}%
     \let\^^\=\@doscaling\scalingskipslist
     \setbox\strutbox=\hbox{\vrule
          height.708\baselineskip depth.292\baselineskip width\z@}}

\def\@getfactor#1#2#3#4{\@@factori=#1 \@@factorii=#2
     \@@factoriii=#3 \@@factoriv=#4}

\def\@getscaling#1{\count@=#1 \advance\count@ by-\@@sizeindex\@@sizeindex=#1
     \ifnum\count@<0
          \let\@mulordiv=\divide
          \let\@divormul=\multiply
          \multiply\count@ by\m@ne
     \else\let\@mulordiv=\multiply
          \let\@divormul=\divide
     \fi
     \edef\@@scratcha{\ifcase\count@                {1}{1}{1}{1}\or
          {1}{7}{23}{3}\or     {2}{5}{3}{1}\or      {9}{89}{13}{1}\or
          {6}{25}{6}{1}\or     {8}{71}{14}{1}\or    {6}{25}{36}{5}\or
          {1}{7}{53}{4}\or     {12}{125}{108}{5}\or {3}{14}{53}{5}\or
          {6}{41}{17}{1}\or    {13}{31}{13}{2}\or   {9}{107}{71}{2}\or
          {11}{139}{124}{3}\or {1}{6}{43}{2}\or     {10}{107}{42}{1}\or
          {1}{5}{43}{2}\or     {5}{69}{65}{1}\or    {11}{97}{91}{2}\fi}%
     \expandafter\@getfactor\@@scratcha}

\def\@doscaling#1{\@mulordiv#1by\@@factori\@divormul#1by\@@factorii
     \@mulordiv#1by\@@factoriii\@divormul#1by\@@factoriv}


\newskip\headskip
\newskip\footskip

\def\typesize=#1pt{\count@=#1 \advance\count@ by-10
     \ifcase\count@
          \@setsizex\or\err@badtypesize\or
          \@setsizexii\or\err@badtypesize\or
          \@setsizexiv
     \else\err@badtypesize
     \fi}

\def\@setsizex{\getixpt
     \def\subsubscriptfonts{\vpt}%
          \def\subsubscriptsize{\vpt\@parameterinit{-8}}%
     \def\subscriptfonts{\viipt}\def\subscriptsize{\viipt\@parameterinit{-4}}%
     \def\footnotefonts{\viiipt}\def\footnotesize{\viiipt\@parameterinit{-2}}%
     \def\smallfonts{\ixpt}\def\smallsize{\ixpt\@parameterinit{-1}}%
     \def\normalfonts{\xpt}\def\normalsize{\xpt\@parameterinit{0}}%
     \def\bigfonts{\xiipt}\def\bigsize{\xiipt\@parameterinit{2}}%
     \def\Bigfonts{\xivpt}\def\Bigsize{\xivpt\@parameterinit{4}}%
     \def\biggfonts{\xviipt}\def\biggsize{\xviipt\@parameterinit{6}}%
     \def\Biggfonts{\xxipt}\def\Biggsize{\xxipt\@parameterinit{8}}%
     \def\tinyfonts{\vpt}\def\tinysize{\vpt\@parameterinit{-8}}%
     \def\HUGEFONTS{\xxvpt}\def\HUGESIZE{\xxvpt\@parameterinit{10}}%
     \normalsize\fixedskipslist}

\def\@setsizexii{\getxipt
     \def\subsubscriptfonts{\vipt}%
          \def\subsubscriptsize{\vipt\@parameterinit{-6}}%
     \def\subscriptfonts{\viiipt}%
          \def\subscriptsize{\viiipt\@parameterinit{-2}}%
     \def\footnotefonts{\xpt}\def\footnotesize{\xpt\@parameterinit{0}}%
     \def\smallfonts{\xipt}\def\smallsize{\xipt\@parameterinit{1}}%
     \def\normalfonts{\xiipt}\def\normalsize{\xiipt\@parameterinit{2}}%
     \def\bigfonts{\xivpt}\def\bigsize{\xivpt\@parameterinit{4}}%
     \def\Bigfonts{\xviipt}\def\Bigsize{\xviipt\@parameterinit{6}}%
     \def\biggfonts{\xxipt}\def\biggsize{\xxipt\@parameterinit{8}}%
     \def\Biggfonts{\xxvpt}\def\Biggsize{\xxvpt\@parameterinit{10}}%
     \def\tinyfonts{\vpt}\def\tinysize{\vpt\@parameterinit{-8}}%
     \def\HUGEFONTS{\xxvpt}\def\HUGESIZE{\xxvpt\@parameterinit{10}}%
     \normalsize\fixedskipslist}

\def\@setsizexiv{\getxiiipt
     \def\subsubscriptfonts{\viipt}%
          \def\subsubscriptsize{\viipt\@parameterinit{-4}}%
     \def\subscriptfonts{\xpt}\def\subscriptsize{\xpt\@parameterinit{0}}%
     \def\footnotefonts{\xiipt}\def\footnotesize{\xiipt\@parameterinit{2}}%
     \def\smallfonts{\xiiipt}\def\smallsize{\xiiipt\@parameterinit{3}}%
     \def\normalfonts{\xivpt}\def\normalsize{\xivpt\@parameterinit{4}}%
     \def\bigfonts{\xviipt}\def\bigsize{\xviipt\@parameterinit{6}}%
     \def\Bigfonts{\xxipt}\def\Bigsize{\xxipt\@parameterinit{8}}%
     \def\biggfonts{\xxvpt}\def\biggsize{\xxvpt\@parameterinit{10}}%
     \def\Biggfonts{\err@sizetoolarge\Biggfonts\HUGEFONTS}%
          \def\Biggsize{\err@sizetoolarge\Biggsize\HUGESIZE}%
     \def\tinyfonts{\vpt}\def\tinysize{\vpt\@parameterinit{-8}}%
     \def\HUGEFONTS{\xxvpt}\def\HUGESIZE{\xxvpt\@parameterinit{10}}%
     \normalsize\fixedskipslist}

\def\subsubscriptfonts{\vpt} \def\subsubscriptsize{\vpt\@parameterinit{-8}}
\def\subscriptfonts{\viipt}  \def\subscriptsize{\viipt\@parameterinit{-4}}
\def\footnotefonts{\viiipt}  \def\footnotesize{\viiipt\@parameterinit{-2}}
\def\smallfonts{\err@sizenotavailable\smallfonts}
                             \def\smallsize{\ixpt\@parameterinit{-1}}
\def\normalfonts{\xpt}       \def\normalsize{\xpt\@parameterinit{0}}
\def\bigfonts{\xiipt}        \def\bigsize{\xiipt\@parameterinit{2}}
\def\Bigfonts{\xivpt}        \def\Bigsize{\xivpt\@parameterinit{4}}
\def\biggfonts{\xviipt}      \def\biggsize{\xviipt\@parameterinit{6}}
\def\Biggfonts{\xxipt}       \def\Biggsize{\xxipt\@parameterinit{8}}
\def\tinyfonts{\vpt}         \def\tinysize{\vpt\@parameterinit{-8}}
\def\HUGEFONTS{\xxvpt}       \def\HUGESIZE{\xxvpt\@parameterinit{10}}

\message{document layout,}


\newtoks\everyoutput \everyoutput={}
\newdimen\depthofpage
\newcount\pagenum \pagenum=0

\newdimen\oddtopmargin  \newdimen\eventopmargin
\newdimen\oddleftmargin \newdimen\evenleftmargin
\newtoks\oddhead        \newtoks\evenhead
\newtoks\oddfoot        \newtoks\evenfoot

\def\topmargin{\afterassignment\@seteventop\oddtopmargin}
\def\leftmargin{\afterassignment\@setevenleft\oddleftmargin}
\def\head{\afterassignment\@setevenhead\oddhead}
\def\foot{\afterassignment\@setevenfoot\oddfoot}

\def\@seteventop{\eventopmargin=\oddtopmargin}
\def\@setevenleft{\evenleftmargin=\oddleftmargin}
\def\@setevenhead{\evenhead=\oddhead}
\def\@setevenfoot{\evenfoot=\oddfoot}

\def\pagenumstyle#1{\@setnumstyle\pagenum{#1}}

\newif\ifdraft
\def\draft{\drafttrue\leftmargin=.5in \overfullrule=5pt }

\def\outputstyle#1{\global\expandafter\let\expandafter
          \@outputstyle\csname#1output\endcsname
     \usename{#1setup}}

\output={\@outputstyle}

\def\normaloutput{\the\everyoutput
     \global\advance\pagenum by\@ne
     \ifodd\pagenum
          \voffset=\oddtopmargin \hoffset=\oddleftmargin
     \else\voffset=\eventopmargin \hoffset=\evenleftmargin
     \fi
     \advance\voffset by-1in  \advance\hoffset by-1in
     \count0=\pagenum
     \expandafter\shipout\pagebox
     \ifnum\outputpenalty>-\@MM\else\dosupereject\fi}

\newdimen\fullhsize
\newbox\leftpage
\newcount\leftpagenum
\newcount\outputpagenum \outputpagenum=0
\let\leftorright=L

\def\twoupoutput{\the\everyoutput
     \global\advance\pagenum by\@ne
     \if L\leftorright
          \global\setbox\leftpage=\leftline{\pagebox}%
          \global\leftpagenum=\pagenum
          \global\let\leftorright=R%
     \else\global\advance\outputpagenum by\@ne
          \ifodd\outputpagenum
               \voffset=\oddtopmargin \hoffset=\oddleftmargin
          \else\voffset=\eventopmargin \hoffset=\evenleftmargin
          \fi
          \advance\voffset by-1in  \advance\hoffset by-1in
          \count0=\leftpagenum \count1=\pagenum
          \shipout\vbox{\hbox to\fullhsize
               {\box\leftpage\hfil\leftline{\pagebox}}}%
          \global\let\leftorright=L%
     \fi
     \ifnum\outputpenalty>-\@MM
     \else\dosupereject
          \if R\leftorright
               \globaldefs=\@ne\head={\hfil}\foot={\hfil}\globaldefs=\z@
               \null\newpage
          \fi
     \fi}

\def\pagebox{\vbox{\makeheadline\pagebody\makefootline}}

\def\makeheadline{%
     \vbox to\z@{\baselinestretch=\@m
          \vskip\topskip\vskip-.708\baselineskip\vskip-\headskip
          \line{\vbox to\ht\strutbox{}%
               \ifodd\pagenum\the\oddhead\else\the\evenhead\fi}%
          \vss}%
     \nointerlineskip}

\def\pagebody{\vbox to\vsize{%
     \boxmaxdepth\maxdepth
     \ifvoid\topins\else\unvbox\topins\fi
     \depthofpage=\dp255
     \unvbox255
     \ifraggedbottom\kern-\depthofpage\vfil\fi
     \ifvoid\footins
     \else\vskip\skip\footins
          \footnoterule
          \unvbox\footins
          \vskip-\footnoteskip
     \fi}}

\def\makefootline{\baselineskip=\footskip
     \line{\ifodd\pagenum\the\oddfoot\else\the\evenfoot\fi}}


\newskip\abovechapterskip
\newskip\belowchapterskip
\newskip\abovesectionskip
\newskip\belowsectionskip
\newskip\abovesubsectionskip
\newskip\belowsubsectionskip

\def\chapterstyle#1{\global\expandafter\let\expandafter\@chapterstyle
     \csname#1text\endcsname}
\def\sectionstyle#1{\global\expandafter\let\expandafter\@sectionstyle
     \csname#1text\endcsname}
\def\subsectionstyle#1{\global\expandafter\let\expandafter\@subsectionstyle
     \csname#1text\endcsname}

\def\chapter#1{%
     \ifdim\lastskip=17sp \else\chapterbreak\vskip\abovechapterskip\fi
     \@chapterstyle{\ifblank\chapternumstyle\then
          \else\newchapternum=\next\chapternumformat\ \fi#1}%
     \nobreak\vskip\belowchapterskip\vskip17sp }

\def\section#1{%
     \ifdim\lastskip=17sp \else\sectionbreak\vskip\abovesectionskip\fi
     \@sectionstyle{\ifblank\sectionnumstyle\then
          \else\newsectionnum=\next\sectionnumformat\ \fi#1}%
     \nobreak\vskip\belowsectionskip\vskip17sp }

\def\subsection#1{%
     \ifdim\lastskip=17sp \else\subsectionbreak\vskip\abovesubsectionskip\fi
     \@subsectionstyle{\ifblank\subsectionnumstyle\then
          \else\newsubsectionnum=\next\subsectionnumformat\ \fi#1}%
     \nobreak\vskip\belowsubsectionskip\vskip17sp }


\let\TeXunderline=\underline
\let\TeXoverline=\overline
\def\underline#1{\relax\ifmmode\TeXunderline{#1}\else
     $\TeXunderline{\hbox{#1}}$\fi}
\def\overline#1{\relax\ifmmode\TeXoverline{#1}\else
     $\TeXoverline{\hbox{#1}}$\fi}

\def\baselinestretch{\afterassignment\@baselinestretch\count@}
\def\@baselinestretch{\baselineskip=\normalbaselineskip
     \divide\baselineskip by\@m\baselineskip=\count@\baselineskip
     \setbox\strutbox=\hbox{\vrule
          height.708\baselineskip depth.292\baselineskip width\z@}%
     \bigskipamount=\the\baselineskip
          plus.25\baselineskip minus.25\baselineskip
     \medskipamount=.5\baselineskip
          plus.125\baselineskip minus.125\baselineskip
     \smallskipamount=.25\baselineskip
          plus.0625\baselineskip minus.0625\baselineskip}

\def\\{\ifhmode\ifnum\lastpenalty=-\@M\else\hfil\penalty-\@M\fi\fi
     \ignorespaces}
\def\newpage{\vfil\break}

\def\lefttext#1{\par{\@text\leftskip=\z@\rightskip=\centering
     \noindent#1\par}}
\def\righttext#1{\par{\@text\leftskip=\centering\rightskip=\z@
     \noindent#1\par}}
\def\centertext#1{\par{\@text\leftskip=\centering\rightskip=\centering
     \noindent#1\par}}
\def\@text{\parindent=\z@ \parfillskip=\z@ \everypar={}%
     \spaceskip=.3333em \xspaceskip=.5em
     \def\\{\ifhmode\ifnum\lastpenalty=-\@M\else\penalty-\@M\fi\fi
          \ignorespaces}}

\def\beginleft{\par\@text\leftskip=\z@ \rightskip=\centering}
     
\def\beginright{\par\@text\leftskip=\centering\rightskip=\z@ }
     
\def\begincenter{\par\@text\leftskip=\centering\rightskip=\centering}

\def\beginnarrow{\defaultoption[\parindent]\@beginnarrow}
\def\@beginnarrow[#1]{\par\advance\leftskip by#1\advance\rightskip by#1}

\begingroup
\catcode`\[=1 \catcode`\{=11 \gdef\beginignore[\endgroup\bgroup
     \catcode`\e=0 \catcode`\\=12 \catcode`\{=11 \catcode`\f=12 \let\or=\relax
     \let\nd{ignor=\fi \let\}=\egroup
     \iffalse}
\endgroup

\long\def\marginnote#1{\leavevmode
     \edef\@marginsf{\spacefactor=\the\spacefactor\relax}%
     \ifdraft\strut\vadjust{%
          \hbox to\z@{\hskip\hsize\hskip.1in
               \vbox to\z@{\vskip-\dp\strutbox
                    \marginnoteformat
                    \vskip-\ht\strutbox
                    \noindent\strut#1\par
                    \vss}%
               \hss}}%
     \fi
     \@marginsf}


\newtoks\everybye \everybye={\par\vfil}
\outer\def\bye{\the\everybye
     \footnotecheck
     \prelabelcheck
     \streamcheck
     \supereject
     \TeXend}

\message{footnotes,}

\newcount\footnotenum \footnotenum=0
\newskip\footnoteskip
\let\@footnotelist=\empty

\def\footnotenumstyle#1{\@setnumstyle\footnotenum{#1}%
     \useafter\ifx{@footnotenumstyle}\symbols
          \global\let\@footup=\empty
     \else\global\let\@footup=\markup
     \fi}

\def\footnote{\footnotecheck\defaultoption[]\@footnote}
\def\@footnote[#1]{\@footnotemark[#1]\@footnotetext}

\def\footnotemark{\defaultoption[]\@footnotemark}
\def\@footnotemark[#1]{\let\@footsf=\empty
     \ifhmode\edef\@footsf{\spacefactor=\the\spacefactor\relax}\/\fi
     \ifnoarg#1\then
          \global\advance\footnotenum by\@ne
          \@footup{\footnotenumformat}%
          \edef\@@foota{\footnotenum=\the\footnotenum\relax}%
          \expandafter\additemR\expandafter\@footup\expandafter
               {\@@foota\footnotenumformat}\to\@footnotelist
          \global\let\@footnotelist=\@footnotelist
     \else\markup{#1}%
          \additemR\markup{#1}\to\@footnotelist
          \global\let\@footnotelist=\@footnotelist
     \fi
     \@footsf}

\def\footnotetext{%
     \ifx\@footnotelist\empty\err@extrafootnotetext\else\@footnotetext\fi}
\def\@footnotetext{%
     \getitemL\@footnotelist\to\@@foota
     \global\let\@footnotelist=\@footnotelist
     \insert\footins\bgroup
     \footnoteformat
     \splittopskip=\ht\strutbox\splitmaxdepth=\dp\strutbox
     \interlinepenalty=\interfootnotelinepenalty\floatingpenalty=\@MM
     \noindent\llap{\@@foota}\strut
     \bgroup\aftergroup\@footnoteend
     \let\@@scratcha=}
\def\@footnoteend{\strut\par\vskip\footnoteskip\egroup}

\def\footnoterule{\normalfonts
     \kern-.3em \hrule width2in height.04em \kern .26em }

\def\footnotecheck{%
     \ifx\@footnotelist\empty
     \else\err@extrafootnotemark
          \global\let\@footnotelist=\empty
     \fi}

\message{labels,}

\let\@@labeldef=\xdef
\newif\if@labelfile
\newwrite\@labelfile
\let\@prelabellist=\empty

\def\label#1#2{\trim#1\to\@@labarg\edef\@@labtext{#2}%
     \edef\@@labname{lab@\@@labarg}%
     \useafter\ifundefined\@@labname\then\else\@yeslab\fi
     \useafter\@@labeldef\@@labname{#2}%
     \ifstreaming
          \expandafter\toks@\expandafter\expandafter\expandafter
               {\csname\@@labname\endcsname}%
          \immediate\write\streamout{\noexpand\label{\@@labarg}{\the\toks@}}%
     \fi}
\def\@yeslab{%
     \useafter\ifundefined{if\@@labname}\then
          \err@labelredef\@@labarg
     \else\useif{if\@@labname}\then
               \err@labelredef\@@labarg
          \else\global\usename{\@@labname true}%
               \useafter\ifundefined{pre\@@labname}\then
               \else\useafter\ifx{pre\@@labname}\@@labtext
                    \else\err@badlabelmatch\@@labarg
                    \fi
               \fi
               \if@labelfile
               \else\global\@labelfiletrue
                    \immediate\write\sixt@@n{--> Creating file \jobname.lab}%
                    \immediate\openout\@labelfile=\jobname.lab
               \fi
               \immediate\write\@labelfile
                    {\noexpand\prelabel{\@@labarg}{\@@labtext}}%
          \fi
     \fi}

\def\putlab#1{\trim#1\to\@@labarg\edef\@@labname{lab@\@@labarg}%
     \useafter\ifundefined\@@labname\then\@nolab\else\usename\@@labname\fi}
\def\@nolab{%
     \useafter\ifundefined{pre\@@labname}\then
          \undefinedlabelformat
          \err@needlabel\@@labarg
          \useafter\xdef\@@labname{\undefinedlabelformat}%
     \else\usename{pre\@@labname}%
          \useafter\xdef\@@labname{\usename{pre\@@labname}}%
     \fi
     \useafter\newif{if\@@labname}%
     \expandafter\additemR\@@labarg\to\@prelabellist}

\def\prelabel#1{\useafter\gdef{prelab@#1}}

\def\ifundefinedlabel#1\then{%
     \expandafter\ifx\csname lab@#1\endcsname\relax}
\def\useiflab#1\then{\csname iflab@#1\endcsname}

\def\prelabelcheck{{%
     \def\^^\##1{\useiflab{##1}\then\else\err@undefinedlabel{##1}\fi}%
     \@prelabellist}}

\message{equation numbering,}

\newcount\chapternum
\newcount\sectionnum
\newcount\subsectionnum
\newcount\equationnum
\newcount\subequationnum
\newcount\figurenum
\newcount\subfigurenum
\newcount\tablenum
\newcount\subtablenum

\newif\if@subeqncount
\newif\if@subfigcount
\newif\if@subtblcount

\def\newchapternum{\newsectionnum=\z@\@resetnum\chapternum}
\def\newsectionnum{\newsubsectionnum=\z@\@resetnum\sectionnum}
\def\newsubsectionnum{\newequationnum=\z@\newfigurenum=\z@\newtablenum=\z@
     \@resetnum\subsectionnum}
\def\newequationnum{\newsubequationnum=\z@\@resetnum\equationnum}
\def\newsubequationnum{\@resetnum\subequationnum}
\def\newfigurenum{\newsubfigurenum=\z@\@resetnum\figurenum}
\def\newsubfigurenum{\@resetnum\subfigurenum}
\def\newtablenum{\newsubtablenum=\z@\@resetnum\tablenum}
\def\newsubtablenum{\@resetnum\subtablenum}

\def\@resetnum#1{\global\advance#1by1 \edef\next{\the#1\relax}\global#1}

\newchapternum=0

\def\chapternumstyle#1{\@setnumstyle\chapternum{#1}}
\def\sectionnumstyle#1{\@setnumstyle\sectionnum{#1}}
\def\subsectionnumstyle#1{\@setnumstyle\subsectionnum{#1}}
\def\equationnumstyle#1{\@setnumstyle\equationnum{#1}}
\def\subequationnumstyle#1{\@setnumstyle\subequationnum{#1}%
     \ifblank\subequationnumstyle\then\global\@subeqncountfalse\fi
     \ignorespaces}
\def\figurenumstyle#1{\@setnumstyle\figurenum{#1}}
\def\subfigurenumstyle#1{\@setnumstyle\subfigurenum{#1}%
     \ifblank\subfigurenumstyle\then\global\@subfigcountfalse\fi
     \ignorespaces}
\def\tablenumstyle#1{\@setnumstyle\tablenum{#1}}
\def\subtablenumstyle#1{\@setnumstyle\subtablenum{#1}%
     \ifblank\subtablenumstyle\then\global\@subtblcountfalse\fi
     \ignorespaces}

\def\eqnlabel#1{%
     \if@subeqncount
          \newsubequationnum=\next
     \else\newequationnum=\next
          \ifblank\subequationnumstyle\then
          \else\global\@subeqncounttrue
               \newsubequationnum=\@ne
          \fi
     \fi
     \label{#1}{\puteqnformat}(\puteqn{#1})%
     \ifdraft\rlap{\hskip.1in{\tt#1}}\fi}

\let\puteqn=\putlab

\def\equation#1#2{\useafter\gdef{eqn@#1}{#2\eqno\eqnlabel{#1}}}
\def\Equation#1{\useafter\gdef{eqn@#1}}

\def\putequation#1{\useafter\ifundefined{eqn@#1}\then
     \err@undefinedeqn{#1}\else\usename{eqn@#1}\fi}

\def\eqnseriesstyle#1{\gdef\@eqnseriesstyle{#1}}
\def\begineqnseries{\subequationnumstyle{\@eqnseriesstyle}%
     \defaultoption[]\@begineqnseries}
\def\@begineqnseries[#1]{\edef\@@eqnname{#1}}
\def\endeqnseries{\subequationnumstyle{blank}%
     \expandafter\ifnoarg\@@eqnname\then
     \else\label\@@eqnname{\puteqnformat}%
     \fi
     \aftergroup\ignorespaces}

\def\figlabel#1{%
     \if@subfigcount
          \newsubfigurenum=\next
     \else\newfigurenum=\next
          \ifblank\subfigurenumstyle\then
          \else\global\@subfigcounttrue
               \newsubfigurenum=\@ne
          \fi
     \fi
     \label{#1}{\putfigformat}\putfig{#1}%
     {\def\marginnoteformat{\tt}\marginnote{#1}}}

\let\putfig=\putlab

\def\figseriesstyle#1{\gdef\@figseriesstyle{#1}}
\def\beginfigseries{\subfigurenumstyle{\@figseriesstyle}%
     \defaultoption[]\@beginfigseries}
\def\@beginfigseries[#1]{\edef\@@figname{#1}}
\def\endfigseries{\subfigurenumstyle{blank}%
     \expandafter\ifnoarg\@@figname\then
     \else\label\@@figname{\putfigformat}%
     \fi
     \aftergroup\ignorespaces}

\def\tbllabel#1{%
     \if@subtblcount
          \newsubtablenum=\next
     \else\newtablenum=\next
          \ifblank\subtablenumstyle\then
          \else\global\@subtblcounttrue
               \newsubtablenum=\@ne
          \fi
     \fi
     \label{#1}{\puttblformat}\puttbl{#1}%
     {\def\marginnoteformat{\tt}\marginnote{#1}}}

\let\puttbl=\putlab

\def\tblseriesstyle#1{\gdef\@tblseriesstyle{#1}}
\def\begintblseries{\subtablenumstyle{\@tblseriesstyle}%
     \defaultoption[]\@begintblseries}
\def\@begintblseries[#1]{\edef\@@tblname{#1}}
\def\endtblseries{\subtablenumstyle{blank}%
     \expandafter\ifnoarg\@@tblname\then
     \else\label\@@tblname{\puttblformat}%
     \fi
     \aftergroup\ignorespaces}

\message{reference numbering,}

\newcount\referencenum \referencenum=0
\newcount\@@prerefcount \@@prerefcount=0
\newcount\@@thisref
\newcount\@@lastref
\newcount\@@loopref
\newcount\@@refseq
\newdimen\refnumindent
\let\@undefreflist=\empty

\def\referencenumstyle#1{\@setnumstyle\referencenum{#1}}

\def\referencestyle#1{\usename{@ref#1}}

\def\@refsequential{%
     \gdef\@refpredef##1{\global\advance\referencenum by\@ne
          \let\^^\=0\label{##1}{\^^\{\the\referencenum}}%
          \useafter\gdef{ref@\the\referencenum}{{##1}{\undefinedlabelformat}}}%
     \gdef\@reference##1##2{%
          \ifundefinedlabel##1\then
          \else\def\^^\####1{\global\@@thisref=####1\relax}\putlab{##1}%
               \useafter\gdef{ref@\the\@@thisref}{{##1}{##2}}%
          \fi}%
     \gdef\endputreferences{%
          \loop\ifnum\@@loopref<\referencenum
                    \advance\@@loopref by\@ne
                    \expandafter\expandafter\expandafter\@printreference
                         \csname ref@\the\@@loopref\endcsname
          \repeat
          \par}}

\def\@refpreordered{%
     \gdef\@refpredef##1{\global\advance\referencenum by\@ne
          \additemR##1\to\@undefreflist}%
     \gdef\@reference##1##2{%
          \ifundefinedlabel##1\then
          \else\global\advance\@@loopref by\@ne
               {\let\^^\=0\label{##1}{\^^\{\the\@@loopref}}}%
               \@printreference{##1}{##2}%
          \fi}
     \gdef\endputreferences{%
          \def\^^\####1{\useiflab{####1}\then
               \else\reference{####1}{\undefinedlabelformat}\fi}%
          \@undefreflist
          \par}}

\def\beginprereferences{\par
     \def\reference##1##2{\global\advance\referencenum by1\@ne
          \let\^^\=0\label{##1}{\^^\{\the\referencenum}}%
          \useafter\gdef{ref@\the\referencenum}{{##1}{##2}}}}
\def\endprereferences{\global\@@prerefcount=\the\referencenum\par}

\def\beginputreferences{\par
     \refnumindent=\z@\@@loopref=\z@
     \loop\ifnum\@@loopref<\referencenum
               \advance\@@loopref by\@ne
               \setbox\z@=\hbox{\referencenum=\@@loopref
                    \referencenumformat\enskip}%
               \ifdim\wd\z@>\refnumindent\refnumindent=\wd\z@\fi
     \repeat
     \putreferenceformat
     \@@loopref=\z@
     \loop\ifnum\@@loopref<\@@prerefcount
               \advance\@@loopref by\@ne
               \expandafter\expandafter\expandafter\@printreference
                    \csname ref@\the\@@loopref\endcsname
     \repeat
     \let\reference=\@reference}

\def\@printreference#1#2{\ifx#2\undefinedlabelformat\err@undefinedref{#1}\fi
     \noindent\ifdraft\rlap{\hskip\hsize\hskip.1in \tt#1}\fi
     \llap{\referencenum=\@@loopref\referencenumformat\enskip}#2\par}

\def\reference#1#2{{\par\refnumindent=\z@\putreferenceformat\noindent#2\par}}

\def\putref#1{\trim#1\to\@@refarg
     \expandafter\ifnoarg\@@refarg\then
          \toks@={\relax}%
     \else\@@lastref=-\@m\def\@@refsep{}\def\@more{\@nextref}%
          \toks@={\@nextref#1,,}%
     \fi\the\toks@}
\def\@nextref#1,{\trim#1\to\@@refarg
     \expandafter\ifnoarg\@@refarg\then
          \let\@more=\relax
     \else\ifundefinedlabel\@@refarg\then
               \expandafter\@refpredef\expandafter{\@@refarg}%
          \fi
          \def\^^\##1{\global\@@thisref=##1\relax}%
          \global\@@thisref=\m@ne
          \setbox\z@=\hbox{\putlab\@@refarg}%
     \fi
     \advance\@@lastref by\@ne
     \ifnum\@@lastref=\@@thisref\advance\@@refseq by\@ne\else\@@refseq=\@ne\fi
     \ifnum\@@lastref<\z@
     \else\ifnum\@@refseq<\thr@@
               \@@refsep\def\@@refsep{,}%
               \ifnum\@@lastref>\z@
                    \advance\@@lastref by\m@ne
                    {\referencenum=\@@lastref\putrefformat}%
               \else\undefinedlabelformat
               \fi
          \else\def\@@refsep{--}%
          \fi
     \fi
     \@@lastref=\@@thisref
     \@more}

\message{streaming,}

\newif\ifstreaming

\def\streamto{\defaultoption[\jobname]\@streamto}
\def\@streamto[#1]{\global\streamingtrue
     \immediate\write\sixt@@n{--> Streaming to #1.str}%
     \newwrite\streamout\immediate\openout\streamout=#1.str }

\def\streamfrom{\defaultoption[\jobname]\@streamfrom}
\def\@streamfrom[#1]{\newread\streamin\openin\streamin=#1.str
     \ifeof\streamin
          \expandafter\err@nostream\expandafter{#1.str}%
     \else\immediate\write\sixt@@n{--> Streaming from #1.str}%
          \let\@@labeldef=\gdef
          \ifstreaming
               \edef\@elc{\endlinechar=\the\endlinechar}%
               \endlinechar=\m@ne
               \loop\read\streamin to\@@scratcha
                    \ifeof\streamin
                         \streamingfalse
                    \else\toks@=\expandafter{\@@scratcha}%
                         \immediate\write\streamout{\the\toks@}%
                    \fi
                    \ifstreaming
               \repeat
               \@elc
               \input #1.str
               \streamingtrue
          \else\input #1.str
          \fi
          \let\@@labeldef=\xdef
     \fi}

\def\streamcheck{\ifstreaming
     \immediate\write\streamout{\pagenum=\the\pagenum}%
     \immediate\write\streamout{\footnotenum=\the\footnotenum}%
     \immediate\write\streamout{\referencenum=\the\referencenum}%
     \immediate\write\streamout{\chapternum=\the\chapternum}%
     \immediate\write\streamout{\sectionnum=\the\sectionnum}%
     \immediate\write\streamout{\subsectionnum=\the\subsectionnum}%
     \immediate\write\streamout{\equationnum=\the\equationnum}%
     \immediate\write\streamout{\subequationnum=\the\subequationnum}%
     \immediate\write\streamout{\figurenum=\the\figurenum}%
     \immediate\write\streamout{\subfigurenum=\the\subfigurenum}%
     \immediate\write\streamout{\tablenum=\the\tablenum}%
     \immediate\write\streamout{\subtablenum=\the\subtablenum}%
     \immediate\closeout\streamout
     \fi}


\def\err@badtypesize{%
     \errhelp={The limited availability of certain fonts requires^^J%
          that the base type size be 10pt, 12pt, or 14pt.^^J}%
     \errmessage{--> Illegal base type size}}

\def\err@badsizechange{\immediate\write\sixt@@n
     {--> Size change not allowed in math mode, ignored}}

\def\err@sizetoolarge#1{\immediate\write\sixt@@n
     {--> \noexpand#1 too big, substituting HUGE}}

\def\err@sizenotavailable#1{\immediate\write\sixt@@n
     {--> Size not available, \noexpand#1 ignored}}

\def\err@fontnotavailable#1{\immediate\write\sixt@@n
     {--> Font not available, \noexpand#1 ignored}}

\def\err@sltoit{\immediate\write\sixt@@n
     {--> Style \noexpand\sl not available, substituting \noexpand\it}%
     \it}

\def\err@bfstobf{\immediate\write\sixt@@n
     {--> Style \noexpand\bfs not available, substituting \noexpand\bf}%
     \bf}

\def\err@badgroup#1#2{%
     \errhelp={The block you have just tried to close was not the one^^J%
          most recently opened.^^J}%
     \errmessage{--> \noexpand\end{#1} doesn't match \noexpand\begin{#2}}}

\def\err@badcountervalue#1{\immediate\write\sixt@@n
     {--> Counter (#1) out of bounds}}

\def\err@extrafootnotemark{\immediate\write\sixt@@n
     {--> \noexpand\footnotemark command
          has no corresponding \noexpand\footnotetext}}

\def\err@extrafootnotetext{%
     \errhelp{You have given a \noexpand\footnotetext command without first
          specifying^^Ja \noexpand\footnotemark.^^J}%
     \errmessage{--> \noexpand\footnotetext command has no corresponding
          \noexpand\footnotemark}}

\def\err@labelredef#1{\immediate\write\sixt@@n
     {--> Label "#1" redefined}}

\def\err@badlabelmatch#1{\immediate\write\sixt@@n
     {--> Definition of label "#1" doesn't match value in \jobname.lab}}

\def\err@needlabel#1{\immediate\write\sixt@@n
     {--> Label "#1" cited before its definition}}

\def\err@undefinedlabel#1{\immediate\write\sixt@@n
     {--> Label "#1" cited but never defined}}

\def\err@undefinedeqn#1{\immediate\write\sixt@@n
     {--> Equation "#1" not defined}}

\def\err@undefinedref#1{\immediate\write\sixt@@n
     {--> Reference "#1" not defined}}

\def\err@nostream#1{%
     \errhelp={You have tried to input a stream file that doesn't exist.^^J}%
     \errmessage{--> Stream file #1 not found}}

\message{jyTeX initialization}

\everyjob{\immediate\write16{--> jyTeX version \fmtversion}%
     \edef\@@jobname{\jobname}%
     \edef\jobname{\@@jobname}%
     \settime
     \openin0=\jobname.lab
     \ifeof0
     \else\closein0
          \immediate\write16{--> Getting labels from file \jobname.lab}%
          \input\jobname.lab
     \fi}


\def\fixedskipslist{%
     \^^\{\topskip}%
     \^^\{\splittopskip}%
     \^^\{\maxdepth}%
     \^^\{\skip\topins}%
     \^^\{\skip\footins}%
     \^^\{\headskip}%
     \^^\{\footskip}}

\def\scalingskipslist{%
     \^^\{\p@renwd}%
     \^^\{\delimitershortfall}%
     \^^\{\nulldelimiterspace}%
     \^^\{\scriptspace}%
     \^^\{\jot}%
     \^^\{\normalbaselineskip}%
     \^^\{\normallineskip}%
     \^^\{\normallineskiplimit}%
     \^^\{\baselineskip}%
     \^^\{\lineskip}%
     \^^\{\lineskiplimit}%
     \^^\{\bigskipamount}%
     \^^\{\medskipamount}%
     \^^\{\smallskipamount}%
     \^^\{\parskip}%
     \^^\{\parindent}%
     \^^\{\abovedisplayskip}%
     \^^\{\belowdisplayskip}%
     \^^\{\abovedisplayshortskip}%
     \^^\{\belowdisplayshortskip}%
     \^^\{\abovechapterskip}%
     \^^\{\belowchapterskip}%
     \^^\{\abovesectionskip}%
     \^^\{\belowsectionskip}%
     \^^\{\abovesubsectionskip}%
     \^^\{\belowsubsectionskip}}


\def\twoupsetup{
     \topmargin=.75in
     \leftmargin=.5in
     \vsize=6.9in
     \hsize=4.75in
     \fullhsize=10in
     \let\draft=\relax}

\outputstyle{normal}                             

\def\marginnoteformat{\subscriptsize             
     \hsize=1in \baselinestretch=1000 \everypar={}%
     \tolerance=5000 \hbadness=5000 \parskip=0pt \parindent=0pt
     \leftskip=0pt \rightskip=0pt \raggedright}

\head={\ifdraft\normalfonts\it\hfil DRAFT\hfil   
     \llap{\number\day\ \monthword\month\ \militarytime}\else\hfil\fi}
\foot={\hfil\normalfonts\numstyle\pagenum\hfil}  

\normalbaselineskip=12pt                         
\normallineskip=0pt                              
\normallineskiplimit=0pt                         
\normalbaselines                                 

\topskip=.85\baselineskip \splittopskip=\topskip \headskip=2\baselineskip
\footskip=\headskip

\pagenumstyle{arabic}                            

\parskip=0pt                                     
\parindent=20pt                                  

\baselinestretch=1000                            


\chapterstyle{left}                              
\chapternumstyle{blank}                          
\def\chapterbreak{\newpage}                      
\abovechapterskip=0pt                            
\belowchapterskip=1.5\baselineskip               
     plus.38\baselineskip minus.38\baselineskip
\def\chapternumformat{\numstyle\chapternum.}     

\sectionstyle{left}                              
\sectionnumstyle{blank}                          
\def\sectionbreak{\vskip0pt plus4\baselineskip\penalty-100
     \vskip0pt plus-4\baselineskip}              
\abovesectionskip=1.5\baselineskip               
     plus.38\baselineskip minus.38\baselineskip
\belowsectionskip=\the\baselineskip              
     plus.25\baselineskip minus.25\baselineskip
\def\sectionnumformat{
     \ifblank\chapternumstyle\then\else\numstyle\chapternum.\fi
     \numstyle\sectionnum.}

\subsectionstyle{left}                           
\subsectionnumstyle{blank}                       
\def\subsectionbreak{\vskip0pt plus4\baselineskip\penalty-100
     \vskip0pt plus-4\baselineskip}              
\abovesubsectionskip=\the\baselineskip           
     plus.25\baselineskip minus.25\baselineskip
\belowsubsectionskip=.75\baselineskip            
     plus.19\baselineskip minus.19\baselineskip
\def\subsectionnumformat{
     \ifblank\chapternumstyle\then\else\numstyle\chapternum.\fi
     \ifblank\sectionnumstyle\then\else\numstyle\sectionnum.\fi
     \numstyle\subsectionnum.}


\footnotenumstyle{symbols}                       
\footnoteskip=0pt                                
\def\footnotenumformat{\numstyle\footnotenum}    
\def\footnoteformat{\footnotesize                
     \everypar={}\parskip=0pt \parfillskip=0pt plus1fil
     \leftskip=1em \rightskip=0pt
     \spaceskip=0pt \xspaceskip=0pt
     \def\\{\ifhmode\ifnum\lastpenalty=-10000
          \else\hfil\penalty-10000 \fi\fi\ignorespaces}}


\def\undefinedlabelformat{$\bullet$}             


\equationnumstyle{arabic}                        
\subequationnumstyle{blank}                      
\figurenumstyle{arabic}                          
\subfigurenumstyle{blank}                        
\tablenumstyle{arabic}                           
\subtablenumstyle{blank}                         

\eqnseriesstyle{alphabetic}                      
\figseriesstyle{alphabetic}                      
\tblseriesstyle{alphabetic}                      

\def\puteqnformat{\hbox{
     \ifblank\chapternumstyle\then\else\numstyle\chapternum.\fi
     \ifblank\sectionnumstyle\then\else\numstyle\sectionnum.\fi
     \ifblank\subsectionnumstyle\then\else\numstyle\subsectionnum.\fi
     \numstyle\equationnum
     \numstyle\subequationnum}}
\def\putfigformat{\hbox{
     \ifblank\chapternumstyle\then\else\numstyle\chapternum.\fi
     \ifblank\sectionnumstyle\then\else\numstyle\sectionnum.\fi
     \ifblank\subsectionnumstyle\then\else\numstyle\subsectionnum.\fi
     \numstyle\figurenum
     \numstyle\subfigurenum}}
\def\puttblformat{\hbox{
     \ifblank\chapternumstyle\then\else\numstyle\chapternum.\fi
     \ifblank\sectionnumstyle\then\else\numstyle\sectionnum.\fi
     \ifblank\subsectionnumstyle\then\else\numstyle\subsectionnum.\fi
     \numstyle\tablenum
     \numstyle\subtablenum}}


\referencestyle{sequential}                      
\referencenumstyle{arabic}                       
\def\putrefformat{\numstyle\referencenum}        
\def\referencenumformat{\numstyle\referencenum.} 
\def\putreferenceformat{
     \everypar={\hangindent=1em \hangafter=1 }%
     \def\\{\hfil\break\null\hskip-1em \ignorespaces}%
     \leftskip=\refnumindent\parindent=0pt \interlinepenalty=1000 }


\normalsize


\def\fmtversion{2.6M (June 1992)}

\catcode`\@=12

\typesize=10pt \magnification=1200 \baselineskip17truept
\footnotenumstyle{arabic} \hsize=6truein\vsize=8.5truein
\input twistopspubnew.lab
\input epsf
\draft
\sectionnumstyle{blank}
\chapternumstyle{blank}
\chapternum=1
\sectionnum=1
\pagenum=0

\def\begintitle{\pagenumstyle{blank}\parindent=0pt
\begin{narrow}[0.4in]}
\def\endtitle{\end{narrow}\newpage\pagenumstyle{arabic}}


\def\beginexercise{\vskip 20truept\parindent=0pt\begin{narrow}[10
truept]}
\def\endexercise{\vskip 10truept\end{narrow}}


\def\eql#1{\eqno\eqnlabel{#1}}
\def\ref{\reference}
\def\peq{\puteqn}
\def\pref{\putref}

\def\mgn{\marginnote}
\def\bex{\begin{exercise}}
\def\eex{\end{exercise}}


\font\open=msbm10 


\def\StretchRtArr#1{{\count255=0\loop\relbar\joinrel\advance\count255 by1
\ifnum\count255<#1\repeat\rightarrow}}
\def\StretchLtArr#1{\,{\leftarrow\!\!\count255=0\loop\relbar
\joinrel\advance\count255 by1\ifnum\count255<#1\repeat}}

\def\StretchLRtArr#1{\,{\leftarrow\!\!\count255=0\loop\relbar\joinrel\advance
\count255 by1\ifnum\count255<#1\repeat\rightarrow\,\,}}

\def\mbox#1{{\leavevmode\hbox{#1}}}

\def\hspace#1{{\phantom{\mbox#1}}}
\def\oZ{\mbox{\open\char90}}

\def\oN{\mbox{\open\char78}}

\def\al{\alpha}
\def\bom{{\bmit\omega}}
\def\be{\beta}

\def\de{\delta}
\def\Ga{\Gamma}

\def\om{\omega}

\def\si{\sigma}

\def\ze{\zeta}

\def\De{\Delta}

\def\caE{{\cal E}}

\def\sc{{\rm sc }}

\def\zf{$\zeta$--function}


\def\frac#1/#2{\leavevmode\kern.1em
\raise.5ex\hbox{\the\scriptfont0 #1}\kern-.1em/\kern-.15em
\lower.25ex\hbox{\the\scriptfont0 #2}}
\def\sfrac#1/#2{\leavevmode\kern.1em
\raise.5ex\hbox{\the\scriptscriptfont0 #1}\kern-.1em/\kern-.15em
\lower.25ex\hbox{\the\scriptscriptfont0 #2}}

\def\gtorder{\mathrel{\raise.3ex\hbox{$>$}\mkern-14mu
             \lower0.6ex\hbox{$\sim$}}}
\def\ltorder{\mathrel{\raise.3ex\hbox{$<$}\mkern-14mu
             \lower0.6ex\hbox{$\sim$}}}

\def\semidirprod{\rlap{\ss C}\raise1pt\hbox{$\mkern.75mu\times$}}
\def\for{\lower6pt\hbox{$\Big|$}}
\def\fish{\kern-.25em{\phantom{abcde}\over \phantom{abcde}}\kern-.25em}


\def\boxit#1{\vbox{\hrule\hbox{\vrule\kern3pt
        \vbox{\kern3pt#1\kern3pt}\kern3pt\vrule}\hrule}}
\def\dalemb#1#2{{\vbox{\hrule height .#2pt
        \hbox{\vrule width.#2pt height#1pt \kern#1pt \vrule
                width.#2pt} \hrule height.#2pt}}}

\def\ol{\overline}
\def\frac#1#2{{{#1}\over{#2}}}

\def\noin{\noindent}

\def\comb#1#2{{\left(#1\atop#2\right)}}

\def\cosec{{\rm cosec\,}}

\def\etc{{\it etc. }}

\def\eg{{\it e.g.}}
\def\ie{{\it i.e. }}
\def\cf{{\it cf }}
\def\pa{\partial}

\def\av#1{\langle#1\rangle} 


\def\3j#1#2#3#4#5#6{\left\lgroup\matrix{#1&#2&#3\cr#4&#5&#6\cr}
\right\rgroup}

\def\m?{\mgn{?}}

\def\pa{\partial}

\def\beq{\begin{eqnarray}}
\def\eeq{\end{eqnarray}}


\def\aop#1#2#3{{\it Ann. Phys.} {\bf {#1}} ({#2}) #3}

\def\cmp#1#2#3{{\it Comm. Math. Phys.} {\bf {#1}} ({#2}) #3}

\def\jmp#1#2#3{{\it J. Math. Phys.} {\bf {#1}} ({#2}) #3}
\def\jpa#1#2#3{{\it J. Phys.} {\bf A{#1}} ({#2}) #3}

\def\np#1#2#3{{\it Nucl. Phys.} {\bf B{#1}} ({#2}) #3}

\def\pr#1#2#3{{\it Phys. Rev.} {\bf {#1}} ({#2}) #3}

\def\prB#1#2#3{{\it Phys. Rev.} {\bf B{#1}} ({#2}) #3}
\def\prD#1#2#3{{\it Phys. Rev.} {\bf D{#1}} ({#2}) #3}
\def\prl#1#2#3{{\it Phys. Rev. Lett.} {\bf #1} ({#2}) #3}

\def\am#1#2#3{{\it Acta Mathematica} {\bf {#1}} ({#2}) #3}

\def\jpamt#1#2#3{{\it J. Phys.A:Math.Theor.} {\bf{#1}} ({#2}) #3}
\def\jram#1#2#3{{\it J. f. reine u. Angew. Math.} {\bf {#1}} ({#2}) #3}

\def\mz#1#2#3{{\it Math. Zeit.} {\bf {#1}} ({#2}) #3}

\def\plb#1#2#3{{\it Phys. Letts.} {\bf {B#1}} ({#2}) #3}

\def\qjm#1#2#3{{\it Quart. J. Math.} {\bf {#1}} ({#2}) #3}

\begin{title}
\vglue 0.5truein
\vskip15truept
\centertext {\Bigfonts \bf Conformal weights of charged R\'enyi entropy} \vskip7truept
\vskip10truept\centertext{\Bigfonts \bf  twist operators for free scalar fields} \vskip17truept
\centertext{\Bigfonts \bf  in arbitrary dimensions}
 \vskip 20truept
\centertext{J.S.Dowker\footnote{ dowker@man.ac.uk;  dowkeruk@yahoo.co.uk}} \vskip
7truept \centertext{\it Theory Group,} \centertext{\it School of Physics and Astronomy,}
\centertext{\it The University of Manchester,} \centertext{\it Manchester, England} \vskip
7truept \centertext{}

\vskip 7truept

\vskip40truept
\begin{narrow}
I compute the conformal weights of the twist operators of free scalar fields for charged
R\'enyi entropy in both odd and even dimensions. Explicit expressions can be found, in odd
dimensions as a function of the chemical potential in the absence of a conical singularity
and thence by images for all integer coverings. This method, developed some time ago, is
equivalent, in results, to the replica technique. A review is given.

The same method applies for even dimensions but a general form is more immediately
available. For no chemical potential, the closed form in the covering order is written in an
alternative way related to old trigonometric sums. Some derivatives are obtained.

An analytical proof is given of a  conjecture made by Bueno, Myers and Witczac--Krempa
regarding the relation between the conformal weights and a corner coefficient (a universal
quantity) in the R\'enyi entropy.

\end{narrow}
\vskip 5truept
\vskip 60truept
\vfil
\end{title}
\pagenum=0
\newpage

\section{\bf 1. Introduction and summary.}

The replica method is  a popular technique in the calculation of entanglement and R\'enyi
entropies which have proved to be valuable probes of aspects of condensed matter physics
and quantum field theory. Relevant references are too many to list and I mention
[\pref{CandH}] and [\pref{CandC}] as representative.

A conformal field--theoretic way of describing the ensuing cut structure involves the notion
of `twist operators', \eg\ [\pref{Hung}], and their conformal weights. These have some
intriguing general properties which have been tested in specific situations, the simplest of
which is free--field theory. My rather restricted aim in the present work is to give some
specific but explicit computations in this area which might prove useful  for comparison or
checking purposes. The quantity I will study particularly is the conformal weight of free
scalars for charged R\' enyi entropy using a method that allows an {\it arbitrary}
dimension, $d$, to be treated easily. $d=3$ is the only odd case treated in detail up to
now. The technique I employ leads to  a proof (section 4) of an interesting conjecture
which relates the conformal weight to a corner coefficient arising in the expansion of
R\'enyi entropy, [\pref{BMW}].

In section 2, I set up some general formulae, transforming the problem to one in a
Euclideanised cosmic string  manifold (\ie flat space with a single conical singularity).
Thereby, in section 3, by putting together some previously derived results, I calculate, for
all odd $d$, the R\'enyi conformal weight, $h_1$, in the absence of the conical singularity
but with a flux along the cone axis which can be interpreted as a Euclidean chemical
potential. An existing image technique (equivalent to a replica relation) then allows the
general R\'enyi weights, $h_n$ ($n\in\oN$), to be found as trigonometric sums. The zero
chemical potential values agree with those that already exist for $d=3$. Sections 7 and 8
give the corresponding even $d$ calculation. Some derivatives are also computed. In
section 10, the resulting polynomials in $n$ are related to finite cosec sums whose
evaluation (due to Jeffery in 1864) organise the polynomials in a more expressive way.
Finally, an appendix contains a derivation of the image technique which was just written
down in an earlier publication, [\pref{Dowcascone}].

\section{\bf 2. The conformal weight}

Since my object is a mere technical evaluation, for rapidity I present the basic equations
as given in Belin {\it et al}, [\pref{B}], without derivations or much explanation.

The conformal weight, $h_n$, of a spherical twist operator is derived in [\pref{B}] using
conformal transformation giving the result (I use their notation initially), in terms of the
energy density, $\caE(T,\mu)$,  on the $d$--dimensional hyperbolic cylinder,
   $$
   h_n(\mu)={2\pi n\over d-1}\big(\caE(T_0,\mu=0)-\caE(T_0/n,\mu)\big)
   \eql{h}
   $$
where $T_0=1/2\pi$ (I  have set the radius $R$ to unity), $\mu$ is the chemical potential
and $n$ is the order of the replica covering.

One way of calculating $\caE$ is to make a further conformal transformation to a flat conical
space \ie to a Euclideanised cosmic string space. This yields the relation (in odd dimensions),
[\pref{Dowconearb}],
  $$
   \caE=-r^d\,(d-1)\,\langle T_{zz}\rangle\,,
   \eql{relnn}
  $$
with the string metric written as
$$
  ds^2=dr^2+r^2 d\phi^2+ d{\bf z.}d{\bf z}\,,
  $$
the angular time point $\phi$ being identified with $\phi+2\pi/q$. I now use $q=2\pi T$ and
think of $q$ as a real number.

The average on the right--hand side of (\peq{relnn}) is with respect to free (complex)
fields, $\psi$, quasi--periodic under $\phi\to\phi+2\pi/q$ \ie
$\psi(\phi+2\pi/q,\ldots)=e^{2\pi i\de}\,\psi(\phi,\ldots)$.\footnote{ As is, by now,
standard, gauge equivalent are periodic fields with an appropriate vector potential in the
equations of motion. See [\pref{dowaustin}] for an extended discussion and other
references.} As described many times before, [\pref{dowaustin, Dowcone, DandB,
schulman2}], $ \de$ could be regarded as a `magnetic flux' through, or along, the `axis'
of the cone which here is the co--dimensional 2 space coordinatised by ${\bf z}$.
Thermally, $T=q/2\pi$ would be the temperature and $\mu= \de$ a (`Euclidean') chemical
potential, [\pref{Dowconearb,DandC}]. \footnote{ The $\mu$ here equals $\mu_E/2\pi$ of
[\pref{B}]. See section 3. I will often call $\de$ the phase.}

The (complex) scalar vacuum energy momentum tensor has been computed with these
boundary conditions awhile ago in [\pref{Dowcascone}] and, in order to avoid duplication,
I just give the answer,
$$\eqalign{
  \av{T_{zz}}&={1\over\pi}{\Ga(d/2)\over (4\pi r^2)^{d/2}}\,q\,
  \bigg(W_d(q,\de)-{d-2\over d-1}\,W_{d-2}(q,\de)\bigg)\cr
  &\equiv{1\over\pi}{\Ga(d/2)\over (4\pi r^2)^{d/2}}\,q\,Y(q,\de)\cr
  &\equiv E(q,\de)\,,
  }
  \eql{t00}
  $$
which defines $Y(q,\de),\,E(q,\de)$ and where $W_d$, on choice of a particular complex
contour, becomes,
  $$
  W_d(q,\de)=\int_0^\infty {d\tau\over\cosh^d \tau/2}\,
  {\cosh(q(\de-1)\tau\big)
  \sin(\pi q\de)-\cosh(q\de \tau)\sin\big(\pi q(\de-1)\big)\over \cosh q\tau-\cos q\pi}\,.
  \eql{wd}
  $$
Here  $q\le1$ which corresponds to a conical angular excess. The values $q=1/n$,
$n\in\oN$, give a multi--sheeted  integral covering of the plane. Zero flux implies $\de=0$
or, equivalently, $\de=1$. Quantities for $\de$ outside the range $0\le\de\le1$ are
computed by periodicity, $\de+1\equiv\de$.

Although it is possible to work generally, at least for even dimensions, a simplification occurs
on setting $q=1$, \ie on removing the conical singularity. I give some of the details that
were not exposed in [\pref{Dowcascone}]. Thus, from (\peq{wd}),
  $$\eqalign{
  W_d(1,\de)&=2\sin\pi\de\int_0^\infty dy\,{\cosh y(2\de-1)\over\cosh^{d+1} y}\cr
  &=2^d\,\cos\pi\ol\de\,{\Ga\big(\ol d+1+\ol\de\big)\,\Ga\big(\ol d+1-\ol\de\big)\over
  \Ga(d+1)}\cr
  &={2^d\over\Ga(d+1)}{\pi\cos\pi\ol\de\over \sin\big(\pi(\ol\de-
  \ol d)\big)}{\Ga\big(\ol \de+\ol d+1\big)\over\Ga\big(\ol\de-\ol d\big)}\,,
  }
  \eql{wd1}
  $$
which is an interpolation between odd and even dimensions. I have set $d=2\ol d+1$ and
$\ol\de=\de-1/2$ generally, and, in the special case of integer $d$, the ratio of Gamma
functions in (\peq{wd1}) factorises according to,
  $$
     {\Gamma\big({x+\ol d+1}
     \big)\over\Gamma\big(x-\ol d\big)}=(-1)^{[d/2]}\,x^{[d+1]-1}\,,
  $$
in terms of central factorials whose definition I repeat here for convenience,
[\pref{Steffensen}],
  $$\eqalign{
  x^{[2\nu]-1}&=x\big(x^2-1^2\big)\big(x^2-2^2\big)\ldots\big(x^2-(\nu-1)^2\big)\cr
  x^{[2\nu+1]-1}&=\big(x^2-\textstyle{1\over4}\big)\big(x^2-\textstyle{9\over4}\big)
  \ldots\big(x^2-\textstyle{(2\nu-1)^2\over4}\big)\,.
  }
  \eql{cf}
  $$

This dimensional continuation could be carried through into the energy density and the
conformal  weights, but I will not pursue this aspect.

Aiming towards the combination in (\peq{t00}),
$$\eqalign{
 W_{d-2}(1,\de) &=2^{d-2}\,\cos\pi\ol\de\,
 {\Ga\big(\ol d+\ol\de\big)\,\Ga\big(\ol d-\ol\de\big)\over
  \Ga(d-1)}\cr
  }
  \eql{wd2}
  $$
so that
  $$
   W_d(1,\de)= 2^2{(\ol d+\ol\de)(\ol d-\ol\de)\over d(d-1)} \,W_{d-2}(1,\de)\,
   \eql{rec}
  $$
whence the combination
  $$\eqalign{
  Y(1,\de)&={2^2W_{d-2}(1,\de)\over d( d-1)}\bigg(\ol d^2-\ol\de^2-{d
  (d-2)\over4}\bigg)\cr
  &={W_{d-2}(1,\de)\over d( d-1)}\,(1-4\,\ol\de^2)\,.
  }
  \eql{comb}
  $$

Regarding this result, I note that $W_d$ vanishes for zero flux ($\ol\de=1/2$)
transcendentally for odd $d$  because of the cosine factor. The conformal combination,
(\peq{comb}), gives an extra {\it algebraic} zero.

The form, (\peq{comb}), is valid only for the range $0<\de\le1$ (or $-1/2<\ol\de\le1/2)$)
and must be extended beyond these by the periodicity coming from the basic definitions.
So far $d$, if integral, could be either even or odd.

From (\peq{cf}) or, equivalently, by iterating (\peq{rec}) down to either $W_0$ (even
$d$) or to $W_1$ (odd $d$), the form of $Y(1,\de)$ can be found. Then, constructing
$\av{T_{zz}}$ from (\peq{t00}), one finds, for odd $d$, the Plancherel form,
  $$\eqalign{
  r^d\,\av{T_{zz}}&={2^d\over\pi}{\pi\Ga(d/2)\over d(d-1)\Ga(d-1)\,(4\pi)^{d/2}}
  \big(\textstyle{{1\over4}}-\ol\de^2\big)\big((\ol d-1)^2-\ol\de^2\big)
  \ldots \big(1-\ol\de^2\big)\,\ol\de\cot\pi\ol\de\cr
  &={\Ga(d/2)\over \Ga(d+1)\,\pi^{d/2}}\,
  \big(\textstyle{{1\over4}}-\ol\de^2\big)\big((\ol d-1)^2-\ol\de^2\big)
  \ldots \big(1-\ol\de^2\big)\,\ol\de\cot\pi\ol\de\cr
  &={1\over(4\pi)^{\ol d}}{1\over (2\ol d+1)\,\,\Ga\big(\ol d+1\big)}\,
  \big(\textstyle{{1\over4}}-\ol\de^2\big)\big((\ol d-1)^2-\ol\de^2\big)
  \ldots \big(1-\ol\de^2\big)\,\ol\de\cot\pi\ol\de\,,
  }
  \eql{tzz}
  $$
which is the answer written out in [\pref{Dowcascone}]. Even $d$ will be treated later in
sections 7 and 8.

The conformal weight follows from (\peq{h}),  and, now just for odd dimensions and
$n=1$, is, on using (\peq{relnn}),
  $$\eqalign{
  h^{(o)}_{1}(\mu)
  &={1\over2(4\pi)^{\ol d-1}}
  {\big({1\over4}-{\mu'}^2\big)\big[\big((\ol d-1)^2-{\mu'}^2\big)
  \ldots \big(1-{\mu'}^2)\big]\,{\mu'}\cot\pi{\mu'}\over (2\ol d+1)\,\Ga\big(\ol d+1\big)}\,,\cr
  }
  \eql{cw2}
  $$
where $\mu'=\mu-1/2$. (Remember, in my conventions, $\mu=1$ corresponds to no
chemical potential, \ie $\mu'=1/2$.)

A plot of this function, suitably periodised, for $d=3$ was given, essentially, in
[\pref{Dowcascone}] and is contained below ($n=1$) in Fig.1, for present circumstances.

\section{\bf3. Use of images to give higher coverings}

So far, to get explicit expressions, I have been restricted to working at $q=1$, \ie at
$n=1$. I now show how to extend the range to $n\in\oN$ using an image method
described in [\pref{Dowcascone}] which gives a decomposition (or sum rule) for the energy
density, $\av{T_{zz}}$, (as an example)\footnote{ The appendix contains more details.}.
From the connection (\peq{relnn}) this translates to,
  $$
  \caE(q/n,n\de)={1\over n}\sum_{s=0}^{n-1}\caE\big(q,\de+s/n\big)\,,
  \eql{images}
  $$
where $n$ (but not necessarily $q$) is integral. Knowing the result for any $\de$ is crucial
for the applicability of the method.

The notation now is such that the first argument of $\caE$ is $2\pi$ times the inverse of
the cone angle, and the second the phase change on circling {\it this} cone {\it once}.
Thus if $q=1$, the left--hand side refers to an $n$--fold covering of the plane with phase
change $n\de$ and so the phase change on circling through just $2\pi$, \ie around one
leaf of the covering, is $\de$, which is the (Euclidean) chemical potential, $\mu$, denoted
$\mu_E/2\pi$ in [\pref{B}].

When $q=1$, the summands are determined by the explicit expression in (\peq{tzz}).
Hence a closed form exists for the left--hand side \ie for the density for an $n$--fold
covering of flat space. This gives a useful, and more elegant, alternative to a crude
numerical treatment of (\peq{wd}) (which, however, applies for all $q\le1$). In this case,
expressed in terms of the conformal weight, (\peq{images}) gives, in view of (\peq{h}),
  $$
  h^{(o,e)}_n(\mu)=\sum_{s=0}^{n-1} h^{(o,e)}_1\big(\mu+s/n\big)\,,
  \eql{images2}
  $$
valid for odd and even dimensions.

A graph of this for $d=3$ appears in Fig.1. Higher dimensions are easily computed. The
meeting at $\mu=1/2$ of the $n=1$  and $n=2$ curves follows from (\peq{images}) on
setting $n=2$.

It is interesting to note that expressions similar to (\peq{images}) hold for those spectral
quantities which are determined linearly in terms of the Green function, for example the
effective action, the heat--kernel and the local \zf. (See the appendix).

\epsfxsize=5truein \epsfbox{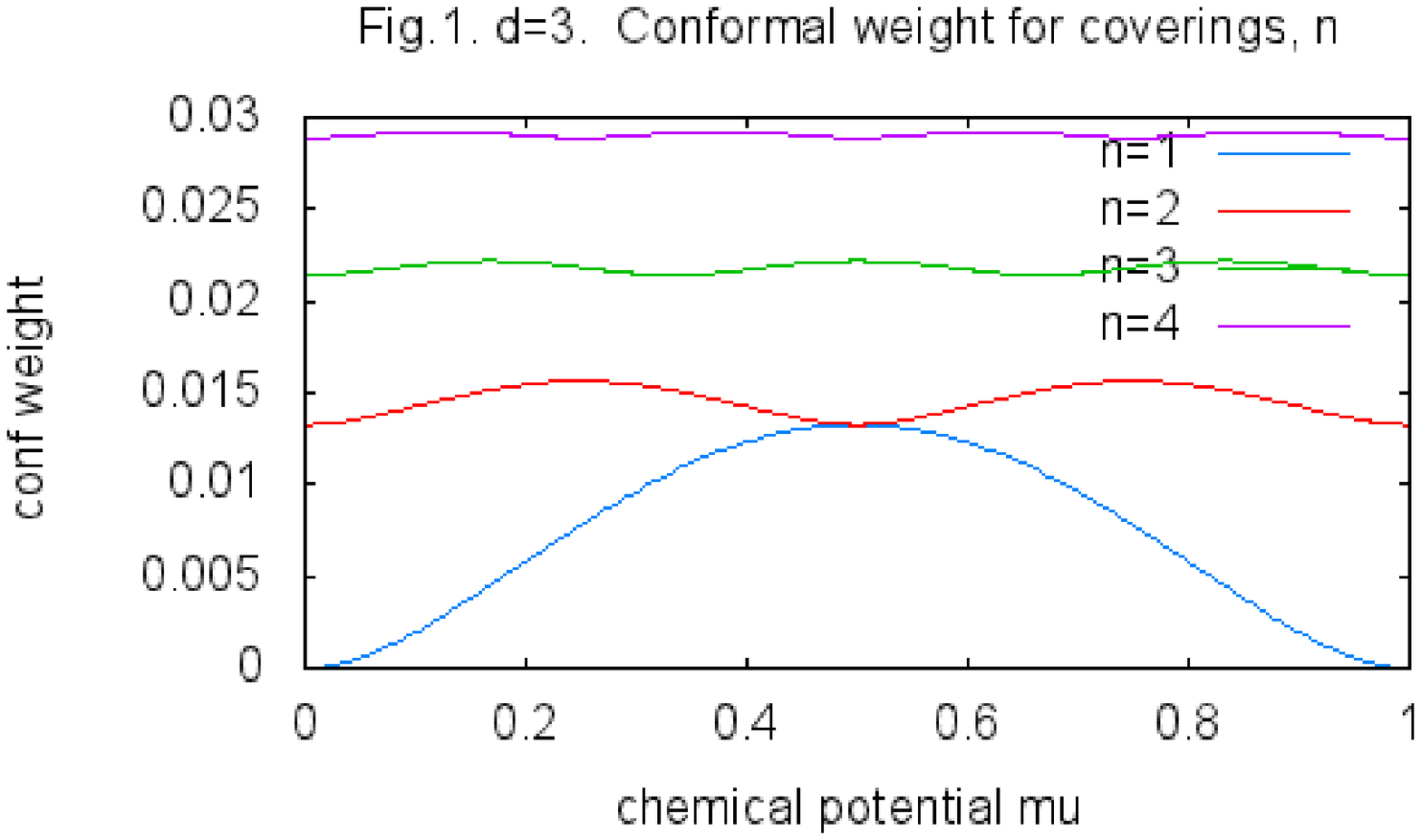}

Exact numerical values  of the $h_n$ (for no chemical potential and $d=3$) were found in
[\pref{BMW}], by a different method, based on earlier expressions. My results for $d=3$
computed using the above formulae agree with those exhibited in Table 1 of this reference
and will not be repeated. As further examples, in five dimensions I find, for $n=5$,
  $$
{3\,(645-83\sqrt5)\sqrt{\sqrt5+5}\over1563\,\, 2^{13/2}\,\pi}\sim
8.3445842623\times 10^{-4}\,,
  $$
and for $n=4$,
  $$
{45\pi+128\over40960\,\pi^2}\sim6.663343844\times 10^{-4}\,.
  $$
Higher dimensions have similar forms which are quickly produced.

\section{\bf4. Proof of a conjecture concerning corner coefficients}

A considerable amount of work has been done to find regularisation independent
(universal) constants that appear during the calculation of entanglement and R\'enyi
entropies. One such constant, $\si_n$, is associated with the existence of a corner, or
vertex, singularity in the entangling surface. The subscript $n\in\oN$ is the order of the
replica covering used to compute the entropies. For the entanglement entropy the limit
$n\to1$ is required but the general case is relevant for R\'{e}nyi entropies.

Casini and Huerta, [\pref{CandH}], gave an integral form for $\si_n$ for free scalar fields
in three dimensions. Motivated by some relations between $\si_n$ and the central charges
conjectured and calculated by Bueno, Myers and Witczak--Krempa, [\pref{BMW2}], Elvang
and Hadjiantonis, [\pref{EandH}], recently have remarkably been able to reduce the rather
complicated integral to a finite sum, which can be evaluated for a given $n$.

Lately, Bueno, Myers and Witczak--Krempa, [\pref{BMW}], have used these results to
make, and verify, the further conjecture that $\si_n$ is related to the conformal weight,
$h_n$, of twist operators by
  $$
  \si_n={1\over\pi}{h_n\over n-1}\,,
  \eql{conj}
  $$
for any three--dimensional conformal field theory.  For free fields, they checked this
numerically for a large range of $n$, using the finite sum for $\si_n$ and an integral form
for $h_n$, derived earlier, which they also transformed into a (different) finite
trigonometric sum for integral $n$.

I begin the proof by writing down Elvang and Hadjiantonis's expression for $\si_n$,
  $$
  \si_n={1\over24\pi\,n^3(n-1)}\sum_{k=1}^{n-1}k(n-k)(n-2k)\,\tan{k\,\pi\over k}\,,
  \eql{EH}
  $$
which they obtained after considerable manipulation.

I show that the conjecture (\peq{conj}) is a simple, perhaps even trivial,  consequence of
(\peq{EH}) and the results above.

In the absence of a conical singularity ($q=1$), the conformal weight in three dimensions, is
(see (\peq{cw2})),
   $$
   h_1(\ol\de)={1\over6}\big({1\over4}-\ol\de^2\big)\,\ol\de\,\cot \pi\ol\de\,,
   \eql{h1}
   $$
where $\ol\de=\de-1/2$. $\de=1$ gives zero flux.

Next the image decomposition, (\peq{images2}), gives, for no chemical potential,
  $$
h_n=\sum_{s=0}^{n-1} h_1\big({s+1\over n}-{1\over2}\big)\,,
\eql{hn}
  $$
and all that remains is to substitute (\peq{h1}) into (\peq{hn}) which gives,
  $$\eqalign{
h_n&=-{1\over6}\sum_{s=0}^{n-1}\bigg({1\over4}-\big({s+1\over n}-{1\over2}\big)^2\bigg)
\bigg({s+1\over n}-{1\over2}\bigg)\tan {\pi(s+1)\over n}\cr
&=-{1\over6}\sum_{k=1}^{n}\bigg({1\over4}-\big({k\over n}-{1\over2}\big)^2\bigg)
\bigg({k\over n}-{1\over2}\bigg)\tan {\pi k\over n}\cr
&={1\over12n^3}\sum_{k=1}^{n-1}{k}(n-k)
(n-2k)\tan{ \pi k\over n}\,.\cr
}
\eql{hn2}
  $$

Comparing (\peq{EH}) and  (\peq{hn2}) produces an analytical proof of the conjecture,
(\peq{conj}), of Bueno, Myers and Witczak-Krempa, [\pref{BMW}], as promised. A factor
of 2 occurs because my analysis uses complex fields.

A similar conclusion can be reached for fermion fields using the fermion image relation
outlined in the appendix.

\section{\bf5. Higher dimensions}
It is easy to find the higher dimensional weights, $h_n$, but the corresponding vertex
constants, $\si_n$, are not yet available.

A conjecture, similar to (\peq{conj}), has been made by Bueno and Myers,
[\pref{BandM}]. For example, in five dimensions,
  $$
  \si_n^{(5)}={16\over9}{h_n\over n-1}\,,
  $$
which, using the results above, yields,
  $$
  \si_n^{(5)}={1\over360\pi\,(n-1)\,n^5}\sum_{k=1}^{n-1}k(n-k)(n-2k)(n+2k)(3n-2k)\,
  \tan{\pi\,k\over n}\,,
  $$
as my prediction for the corner contribution in 5 dimensions. Other dimensions are easily
calculated.

\section{\bf6. Derivatives}

Although (\peq{cw2}) is an explicit expression for the conformal weight, $h_1(\mu)$, it is
helpful to evaluate a few lower derivatives with respect to $\mu$  at $\mu=1\equiv0$ in
order to compare, if possible, with other results.

I define $h^{(o)}_{0,p}=\pa_\mu^p\,h^{(o)}_1(\mu)\big|_{\mu=1}$ to agree with the
notation of [\pref{B}]. Because of the double zero at $\mu'=1/2$, the first derivative
$h^{(o)}_{0,1}$ vanishes, in agreement with the remarks in [\pref{B}]. This would not
be so in the non--conformal case. The second derivative also follows easily as the specific
function of dimension,
  $$
  h^{(o)}_{0,2}={\Ga(\ol d+1/2)\,\Ga(\ol d-1/2)\over(4\pi)^{(d-3)/2}}\,.
  $$

I could not find anything in [\pref{B}] with which to compare this quantity directly.

\section{\bf 7. Even dimensions}

I now turn to the easier case of even dimensions and I will, initially, again restrict to the
$q=1$ case \ie no conical singularity, only an Aharonov--Bohm flux. The relevant
expression is given in (\peq{wd1}). The general formulae are the same but now
  $$
  W_0=\sin\pi\de\,\Ga(\de)\Ga(1-\de)=\pi
  $$
and the expressions are entirely algebraic. In place of (\peq{tzz}) one has,
[\pref{Dowcascone}],
  $$\eqalign{
   r^d\,\av{T_{zz}}&={1\over(4\pi)^{\ol d}}{1\over (2\ol d+1)\,\Ga\big(\ol d+1\big)}
  \big(\textstyle{{1\over4}}-\ol\de^2\big)\big[\big({1\over4}-\ol\de^2\big)
  \ldots \big((\ol d-1)^2-\ol\de^2\big)\big]\,,
  }
  \eql{tzz2}
  $$
 where, in the particular case of $\ol d=1/2$, the product in square brackets is empty and there is
only {\it one} zero at $\ol\de=1/2$.

Correspondingly, the conformal weight is,
 $$\eqalign{
  h^{(e)}_{1}(\mu)
  &=
  {(4\pi)^{1-\ol d}\over 2(2\ol d+1)\,\Ga\big(\ol d+1\big)}\textstyle\big({1\over4}-{\mu'}^2\big)
  \big[\big({1\over4}-{\mu'}^2\big)
  \ldots \big((\ol d-1)^2-{\mu'}^2\big)\big]\,.\cr
  }
  \eql{cw3}
  $$

When $\ol d=1/2$, because of the periodicity in $\mu$, the first derivative with respect to
$\mu$ suffers a discontinuity at $\mu=0,1$ as shown in Fig.2. \footnote{ This is the same
as the discontinuity in the second virial coefficient for anyons, $\de$ corresponding to the
statistics determining parameter there, [\pref{Dowstat}].}

The image method used previously to extend $q=1$ to $q=1/n$ applies here as well but,
for even dimensions, it is also possible to give a closed expression for any $q$, \ie any
cone angle, [\pref{Dowcascone}]. This is because the complex contour can be chosen to
encircle the origin and the result of a standard residue calculation is of polynomial form.
Although the analysis is available in the cited references, in order to be self--contained I
present some essentials in section 9. Here I point out that the polynomials are given
explicitly in [\pref{Dowcascone}], which contains a list of conformal conical energy
densities with numerical coefficients. For example, the $d=4$ conformal weight as a
function of the chemical potential and covering order is,
  $$
  h_n^{(4)}={1\over3n^3}\bigg({n^4-1\over45}+{2\over3}\mu^2(\mu-1)^2
  \bigg)\,.
  $$

A systematic way of finding such polynomials is given in section 9. Furthermore, section 10
has another way of organising the polynomials by relating them to trigonometric sums.

Plots are given for $d=2$ and $d=4$ and a few coverings in Figs.2 and 3. \vskip .5in
\epsfxsize=5truein \epsfbox{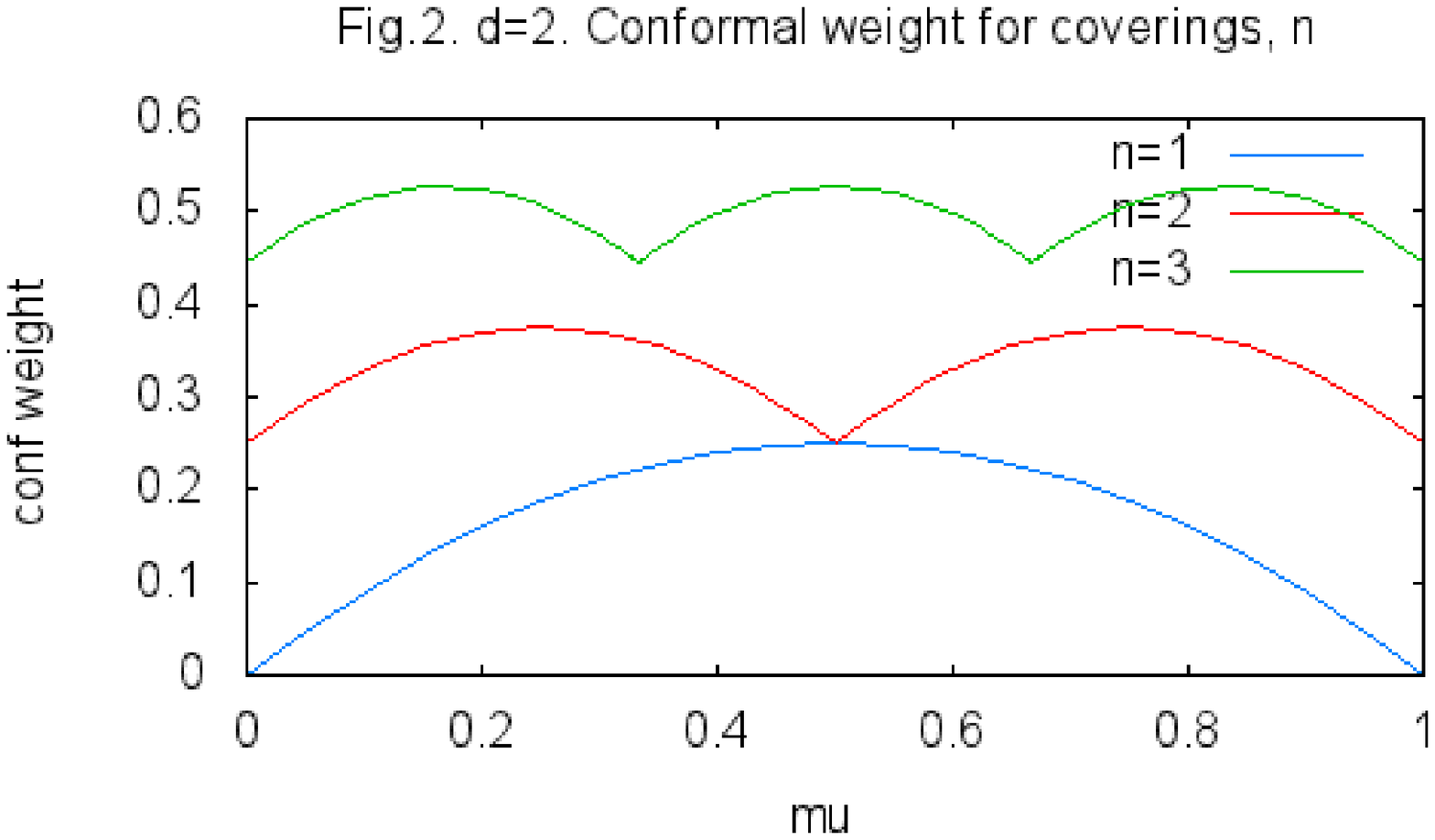}

\newpage

\epsfxsize=5truein \epsfbox{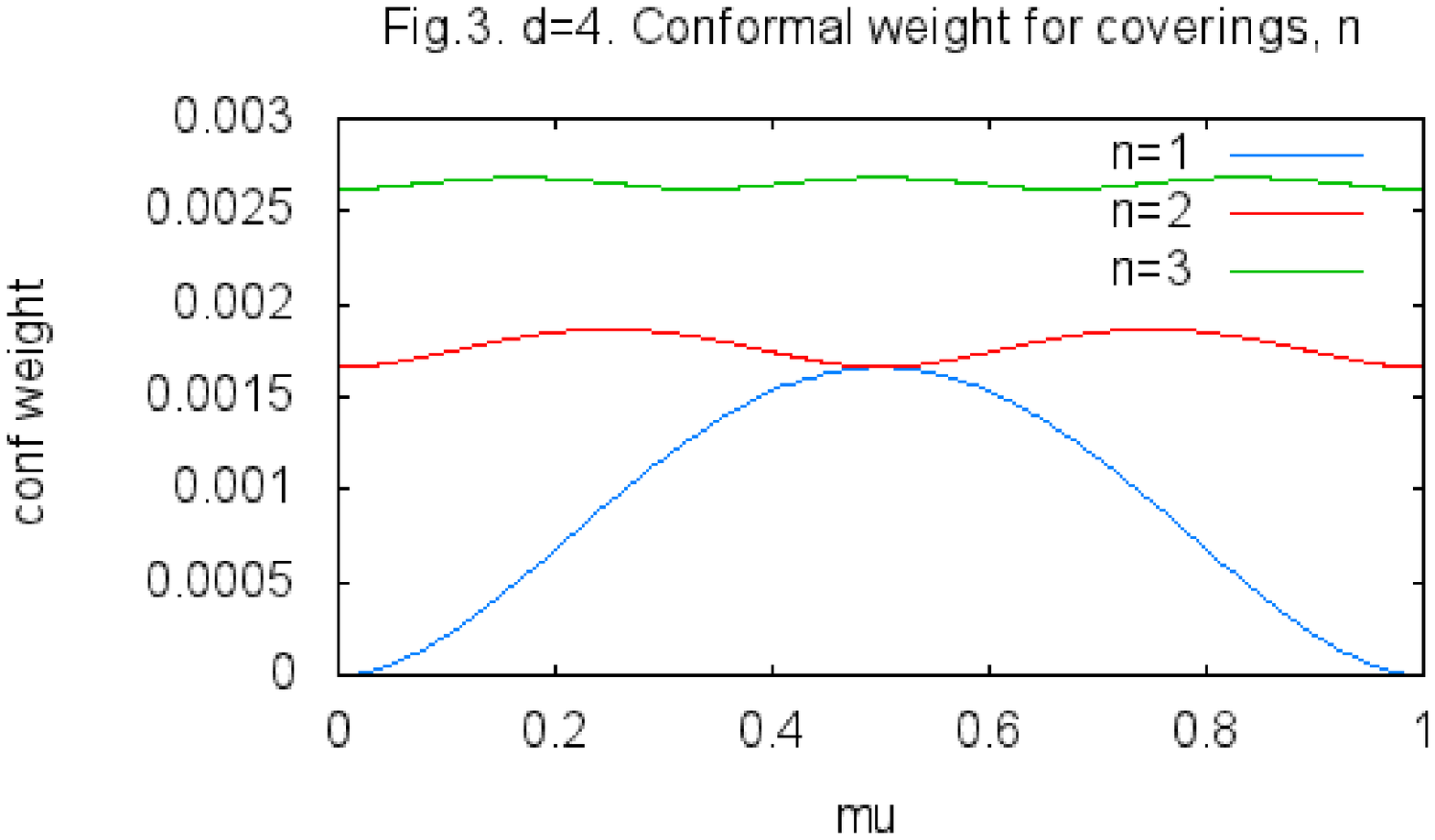}

\section{\bf 8. Singularities. Phase transitions}

The explicit expressions, (\peq{cw2}) and  (\peq{cw3}), show that derivatives of the
plane weights $h^{(o,e)}_1(\mu)$ have discontinuities at $\mu=0,1$ as a consequence of
the imposed periodicity. The first derivative with respect to $\mu$, or $\mu'$, for $d=2$,
the third for $d=3$ and $4$ the fifth for $d=5$ and 6, and so on, as briefly noted in
[\pref{Dowcascone}].

The image expression (\peq{images}) is periodic in $\mu$ with period $1/n$ and shows
that the singular points for the weights of the $n$--fold  cover are at $\mu=s/n$, $s=1$
to $s=n$.  (This is most obvious in Fig.2.)

Such discontinuities in thermal physics usually signal phase transitions. A similar
conclusion in two dimensions is reached by Belin {\it et al}, [\pref{B}], by a different
method. For a free massless fermion, they determine the $n$ positions of the transitions
to be at $\mu=(2k-1)/2n$ to $\mu=(2k+1)/2n$ with $k$ an integer.

\section {\bf 9. Even $d$ again. Derivatives}

By continuing the polar angle into the complex plane, Sommerfeld used a contour method
to find the Green function, $G$, or an equivalent quantity like the diffusion kernel, on a
cone of {\it arbitrary} angle. The method amounts to the re--periodisation of the Green
function without a conical singularity, $G_0$. Conveniently, a contour manipulation
enables $G_0$ to be subtracted from $G$ from the start. This can be thought of as a
renormalisation. One can also allow for a flux along the cone axis by appropriate choice of
$G_0$.

For even dimensions (to which the following analysis is restricted) the relevant massless
$G_0$ possesses only poles and the contour can be deformed into a closed one around a
point which can be taken to be the origin for a particular, inessential, choice of angular
coordinate.

Computation of field theory quantities from $G$, such as vacuum averages, proceeds by
action on this contour integral. A summary of this approach is given in [\pref{Dowstring}].

The method works if a conical part can be displayed in the metric and if $G_0$ is known.
Such is the case for flat space, of course, and gives the cosmic string expressions as used in
this paper, and in [\pref{Dowcascone}].

The vacuum averages of interest here then devolve upon the computation, in even
dimensions, of the quantity, see (\peq{t00}),
  $$
  W_d(q,\de)={i\over4}\oint_C d\al{\cos(q\ol\de\al)\over\sin^d(\al/2)\sin(q\al/2)}
  \eql{wd3}
  $$
where $C$ is a small clockwise\mgn{changed oct 2015} loop around $\al=0$. A closed
form for this can be found in terms of Bernoulli polynomials, as detailed in
[\pref{Dowcascone}], and below. I set, thermally, $\ol\de=\de-1/2=\mu-1/2=\mu'$.

The needed power series expansion of the integrand  in (\peq{wd3}) can be given in terms
of higher Bernoulli polynomials, $B^{(n)}_\nu$. Equivalent to these are the, here more
convenient, higher N\"orlund $D$--polynomials defined, generally, by the central
translation,
  $$
  D^{(n)}_\nu(x\mid{\bom})=2^\nu\,B^{(n)}_\nu
  \big(x+({\textstyle{\sum}}\,\om_i)/2\mid{\bom}\big)
  \eql{DB}
  $$
where ${\bom}$ is an $n$--vector of components $\om_i$. (Refer to N\"orlund,
[\pref{Norlund}], for any unexplained notation). In the case here, $n=d+1$ and
${\bom}=(q,{\bf1}_d)$.

The standard expansions then yield,{\mgn{Checked in dees.wxm}
  $$
  {\cos(q\mu'\al)\over\,\sin^d(\al/2)\sin(q\al/2)}={2^{d+1}\over q\,\al^{d+1}}
  \sum_{\nu=0}^\infty
  {(-1)^\nu\over (2\nu)!}\,\left({\al\over2}\right)^{2\nu}\,
  D^{(d+1)}_{2\nu}\big(q\mu'\mid q,{\bf 1}_d\big)\,.
  \eql{exp}
  $$
Hence, by residues, for even $d$, the exact answer is,\mgn{correct for clockwise loop}
  $$
   W_d(q,\mu)=
  {(-1)^{d/2}\,\pi\over\,d!\,q}D^{(d+1)}_{d}
  \big(q\mu'\mid q,{\bf 1}_d\big)\,.
  \eql{why2}
  $$

Although this can be taken as a general closed form, I will, equivalently, concentrate on
the derivatives with respect to $\mu'$ and $q$ at $\mu'=1/2$ and $q=1$. This will be
sufficient to give the derivatives of $h_n=h_{1/q}$ from (\peq{h}), (\peq{relnn}) and
(\peq{t00}). References [\pref{Hung}] and [\pref{B}] give general properties of these
derivatives so a specific evaluation is useful.

I prefer the derivatives with respect to $q$ rather than $n,=1/q$, as $q\,W_D(q,\mu')$ is
a bipolynomial in $q$ and $\mu'$ and most derivatives become zero after a while. The
derivatives with respect to $n$ can easily be reconstituted.

Pursuing this, I define the derivatives,
  $$\eqalign{
  W_{ab}(d,q,\mu')&\equiv {1\over a!\,b!}\,\pa_{\mu'}^a\,\pa_q^b \,W_d(q,\mu)\cr
  W_{ab}(d)&\equiv  W_{ab}(d,q,\mu')\bigg|_{\mu'=1/2,\,q=1}\,.
  }
  \eql{why}
  $$

If one wishes to display the polynomial content of (\peq{why}) explicitly then there is the
expansion, [\pref{Norlund}] p.162,
  $$
  D^{(d+1)}_{d}
  \big(q\mu'\mid q,{\bf 1}_d\big)=
  \sum_{s=0}^d\comb d s {(2q\mu')}^s\,D_{d-s}^{(d+1)}[q,{\bf 1}_d]\,,
  \eql{bees}
  $$
in terms of higher N\"orlund $D$--{\it numbers}, $D^{(n)}_\nu[*]$, which are zero if
$\nu$ is odd so the sum in (\peq{bees}) is over even integers because, here, $d$ is even.

Furthermore, the $D[*]$ are themselves polynomials in $q$, with,
  $$
  q^{s-1}\,D_{d-s}^{(d+1)}[q,{\bf 1}_d]=
  \sum_{t=0}^{d-s}\comb {d-s}t q^{t+s-1}\,D_t^{(1)}[1]\,D^{(d)}_{d-s-t}[{\bf 1}_d]\,.
  \eql{dexp}
  $$
For shortness, the quantities with all parameters equal to 1 (historically the most studied
case) are denoted by,
  $$
  D^{(d)}_\nu\equiv D^{(d)}_{\nu}[{\bf 1}_d]\,,
  $$
and are sometimes referred to as {\it the} (higher) N\"orlund $D$--numbers. In particular
one sets $D_\nu\equiv D_\nu^{(1)}$. These are the $D$--analogues of the more familiar
Bernoulli numbers and can be computed in similar ways. Then setting $q=1$ in
(\peq{dexp}) gives a recursion for the higher numbers. There are other means of finding
them however.

Simply as a check at this point, I confirm that the previously found product form, \eg\
(\peq{tzz2}), at $q=1$, results from the method of this section. From (\peq{why2}), it is
sufficient to give the polynomials, $D^{(d+1)}_d(x)\equiv D^{(d+1)}_d(x\mid
{\bf1}_{d+1})$. Here $d$ is even.

The result follows from the standard descending factorial form for the $B^{(n)}_{n-1}(x)$
given by N\"orlund, [\pref{Norlund}], eqn, (4) p.186 as a simple consequence of recursion.
The transition to $D$, (\peq{DB}), corresponds to a central translation of $B$, and gives
the (even) central factorial,
  $$\eqalign{
  D^{(d+1)}_d(x)&=2^d\big(x^2-\textstyle{\frac 1 4}\big)\big(x^2-\textstyle{\frac 9 4}\big)
  \ldots\big(x^2-\textstyle{\frac {(d-1)^2} 4}\big)\cr
  }
  \eql{deee}
  $$
which, when used in (\peq{why2}), allows expression (\peq{tzz2}) to be
regained.\footnote{ The coefficients of the expansion (\peq{deee}) in powers of $x^2$
are defined to be the central factorial numbers, or the central differentials of nothing.
Then (\peq{bees}) produces a relation between these quantities and the higher N\"orlund
$D$--numbers. All this is more or less standard but not widely used. Consult \ N\"orlund's
classic text, [\pref{Norlund1}], for some relevant expansions. The expression
(\peq{deee}) also yields a combinatorial form for these $D$--numbers as the product of
the first $d/2$ odd numbers taken a certain number of times.}

The series expressions are easily programmed and I give two examples for the derivatives,
$h^{(e)}_{(a,b)}$, of the conformal weights obtained by combining (\peq{why}),
(\peq{h}), (\peq{relnn}) and (\peq{t00}). I display them without the factor of
$1/(4\pi)^{d/2}$ appearing in (\peq{t00}) and also without the factor of $\pi$ in
(\peq{h}) in order to give simple fractions. For $d=2$,
$$
\matrix{b\backslash a&0&1&2&3\cr{0}&0&-4&-4&0&\cr
{1}&-{4}/{3}&-4&-4&0&\cr
{2}&2/3&0&0&0\cr{3}&-2/3&0&0&0&\cr}\,.
$$
and for $d=4$,
  $$
\matrix{b\backslash a&0&1&2&3&4\cr{0}&0&0&2/3&4/3&2/3\cr
{1}&-4/45&0&2&4&2\cr
{2}&-2/45&0&2&4&2\cr
{3}&-2/45&0&2/3&4/3&2/3\cr{4}&1/45\,&0&0&0&0}\,.
$$
Note that, unless $a=0$, the derivatives terminate beyond $d$ or $d-1$  because of the
polynomial structure.

Other dimensions can easily be computed. Ideally one would like to have the dependence on
dimension displayed explicitly but this seems more difficult.

\begin{ignore}
\section{\it Alternative treatment of residue}

Going back to (\peq{wd3}), this time I  define, for convenience, $\ol W(d,q,\mu')\equiv
W_d(q,{\mu'\over q})$.

This will be sufficient to give the derivatives of $\av {T_{zz}}$ which, from (\peq{t00}),
involve the combination,
  $$
  \pa^b_q \,q\, \ol W(d,q,\mu')\bigg|_{q=1}=\big(\pa^b_q+b\,\pa^{b-1}_q\big)
  \ol W(d,q,\mu')\bigg|_{q=1}\,.
  $$

Pursuing this, I now define ($z=\al/2$),
  $$\eqalign{
  \ol W_{ab}(d)&\equiv \pa_{\mu'}^a\,\pa_q^b \,\ol W(d,q,\mu')\bigg|_{\mu'=1/2,q=1}\cr
  &=i\,2^{b-1}\oint_C dz\,{z^{a+b}\over\sin^d z}\,\pa_z^a\cos z\,\,
  \pa^b_z\,\cosec z\,,
  }
  \eql{why3}
  $$
and, initially, distinguish four cases arising from the choice of $a$ and $b$, both being either
odd or even.

The integral is evaluated by finding the small $z$ expansion of the integrand. This can be
done directly, but it is better to take the derivatives before expanding.

The derivative of $\cos z$ is trivial, giving either a single sine or cosine, but that of $\cosec
z$ is more complicated. One can use the derivative polynomial  for $\cosec$ or, equivalently
of $\sec$. Such an  expansion is given by Stern, [\pref{Stern}]. He has one for even
derivatives, (he actually uses $\sec$),
  $$
  \pa_z^{2\si}\,\cosec z=\cosec z\sum_{k=0}^\si
  u^{(2\si)}_k\cot^{2k}z\,,
  \eql{sterne}
  $$
and one for odd,
  $$\eqalign{
  \pa_z^{2\si+1}\,\cosec z
  &=\cosec z \sum_{k=0}^\si u^{(2\si+1)}_k\cot^{2k+1}z\,.
  }
  \eql{sterno}
  $$

The coefficients satisfy coupled recursions, [\pref{Stern}], but, for my purposes, I need to
convert the $\cot^{2k}$ into (even) powers of $\cosec x$ and write, accordingly,
  $$
  \pa_z^{2\si}\,\cosec z=\sum_{k=0}^\si
  v^{(2\si)}_k\cosec^{2k+1}z\,,
  \eql{sterne2}
  $$
and,
  $$\eqalign{
  \pa_z^{2\si+1}\,\cosec z
  &=\cot z\,
  \sum_{k=0}^\si v^{(2\si+1)}_k\,\cosec^{2k+1}z\,.
  }
  \eql{sterno2}
  $$
I assume that the $u$ and $v$ coefficients are known since they are easily computed.

I consider the four cases in turn and comment on them afterwards.

\noindent First, $a$ and $b$ even, $a=2\rho,\,b=2\si$. The integrand in (\peq{why3})
reads
  $$
   I_1=(-1)^\rho z^{2(\rho+\si)}\cos z\,\sum_{k=0}^\si
   v_k^{(2\si)}\cosec^{2k+1+d}z
  $$
Second, $a=2\rho$ and $b=2\si+1$,\mgn{check}
   $$
   I_2=(-1)^\rho z^{2(\rho+\si)+1}\,\sum_{k=0}^\si
   v_k^{(2\si+1)}\big(\cosec^{2k+d+2}z-\cosec^{2k+d}z\big)
  $$
Third, $a=2\rho+1$ and $b=2\si$,
  $$
   I_3=(-1)^{\rho+1} z^{2(\rho+\si)+1} \,\sum_{k=0}^\si
   v_k^{(2\si)}\cosec^{2k+d}z
  $$
Fourth, $a=2\rho+1$ and $b=2\si+1$,
  $$
   I_4=(-1)^{\rho+1} z^{2(\rho+\si+1)}\cos z\,\sum_{k=0}^\si
   v_k^{(2\si+1)}\cosec^{2k+d+1}z\,.
  $$

One sees that to calculate these expressions, just two integrals are required, of the general
forms,
  $$
J_1(\xi,\eta)=  \oint dz\, {z^{\xi}\over\sin^\eta z},\,\quad
J_2(\xi,\eta)=  \oint dz\, {z^{\xi}\,\cos z\over\sin^\eta z}=
{\xi\over1-\eta}\,J_1(\xi-1,\eta-1)\,,
  $$
and $J_1$ is easily evaluated in terms of the known N\"orlund $D$--numbers, $D$, using
the defining expansion
  $$
  {z^\eta\over\sin^\eta z}=\sum_{\nu=0}^\infty (-1)^\nu\,D^{(\eta)}_{2\nu}
  {z^{2\nu}\over(2\nu)!}\,.
  $$
leading to
  $$
  J_1(\xi,\eta)=-2\pi i (-1)^{\eta-\xi}
  {D^{(\eta)}_{\eta-\xi-1}\over(\eta-\xi-1)!}\,.
  \eql{jay1}
  $$

For the complete quantity (\peq{why3}), therefore,\mgn{signs}
  $$\eqalign{
  \ol W^{(e,e)}_{ab}(d)&=2\pi(a+b)\sum_{k=0}^{b/2}(-1)^k\,v_k^{(b)}\,
  {D^{(2k+d)}_{2k+d-a-b}\over(2k+d)\,(2k+d-a-b)!}\cr
   \ol W^{(e,o)}_{ab}(d)&=2\pi\sum_{k=0}^{(b-1)/2}(-1)^k\,v_k^{(b)}\,
  \bigg({D^{(2k+d+2)}_{2k+d-a-b+1}\over(2k+d-a-b+1)!}-
  {D^{(2k+d)}_{2k+d-a-b-1}\over(2k+d-a-b-1)!}\bigg)\cr
  \ol W^{(o,e)}_{ab}(d)&=2\pi\sum_{k=0}^{b/2}(-1)^k\,v_k^{(b)}\,
  {D^{(2k+d)}_{2k+d-a-b-1}\over(2k+d-a-b-1)!}\cr
  \ol W^{(o,o)}_{ab}(d)&=2\pi(a+b)\sum_{k=0}^{(b-1)/2}(-1)^k\,v_k^{(b)}\,
  {D^{(2k+d)}_{2k+d-a-b}\over(2k+d)\,(2k+d-a-b)!}
  }
  \eql{ws}
  $$

A simple consequence is that $\ol W^{(e,e)}_{ab}$ vanishes if $a>d$,  $\ol W^{(e,o)}$ if
\mgn{check} $a>d$, $\ol W^{(o,e)}$ if $a>d-1$ and $\ol W^{(o,o)}$ if $a>d-1$. This just
reflects the polynomial nature of $\ol W$ in $\mu'$.
\end{ignore}
\section{\it Computation of the $D$-numbers}
As a brief aside, I turn to the purely mathematical topic of the computation of the
N\"orlund $D$--numbers, a well known occupation. Equation (\peq{dexp}) involves the
even variety, $D^{(2n)}_{2n-2l}$.

Probably the most useful expression is the one which comes from the relation with the
central differentials of nothing (or the central factorial numbers). This is.
  $$\eqalign{
  D^{(2n)}_{2n-2l}&=2^{2n-2l}{(2n-2l)!\over 2l\,(2n-1)!}\bigg(D_x^{2l}\prod_{i=0}^{n-1}
  (x^2-i^2)\bigg)\bigg|_{x=0}\cr
  &=2^{2n-2l}{(2n-2l)!\over (2n-1)!}\bigg(D_x^{2l-1}x\,\prod_{i=1}^{n-1}
  (x^2-i^2)\bigg)\bigg|_{x=0}\,
  }
  \eql{don}
  $$
where $D_x=d/dx$. The second expression in (\peq{don}) also results from an application
of the relation between the $D^{(n)}_\nu$ and the generalised Bernoulli polynomials
written in terms of products, [\pref{Norlund}] p.193. It is less convenient than the top line
which has a neater combinatorial form, the limit of the term in brackets being equal to the
product of the squares of the first $n-1$ integers, taken a certain number of times.
Precisely,
  $$\eqalign{
  {D^{2l}\,0^{[2n]}\over (2l)!}&={\rm co}_{2l}\,\prod_{i=0}^{n-1}  (x^2-i^2)=
  {\rm co}_{2l-2}\,\prod_{i=1}^{n-1}  (x^2-i^2)\cr
  &=(-1)^l\,\sum_{{i_1<\ldots <i_{2l}\atop =1}}^{n-1}\,i_1^2\,i_2^2\ldots i_{n-l}^2\,.
  }
  $$
This can be taken as an explicit computational formula for $D^{(2n)}_{2\nu}$, if $\nu<n$,
  $$
  D^{(2n)}_{2n-2l}=(-1)^l\,2^{2n-2l}\comb{2n-1}{2l-1}^{-1}
  \sum_{{i_1<\ldots <i_{n-l}\atop =1}}^{n-1}\,i_1^2\,i_2^2\ldots i_{n-l}^2\,,
  $$
which avoids recursion and Bernoulli numbers.

See also Mansour and Dowker, [\pref{MandD}].

\section{\bf10. Trigonometric sums}

In this final section, I derive an alternative organisation of the polynomials in $n$ which
occur in the  conformal weights in even dimensions, $d=2g$.

A different evaluation of the contour integral (\peq{wd3}) yields the well known twisted
trigonometric image sum, which occurs in many places, (see [\pref{Dow7}]),\mgn{Checked
with (\peq{why2}) in dees.wxm}
  $$
  W_d(q,\de)=-{2\pi \over q}\,\sum_{l=1}^{q-1}\cos(2\pi\de l)\,\cosec^d{\pi l\over q}\,,
  \eql{trigsum}
  $$
where the flux, $\de=r/q$, is rational. I deal mostly with the case of no flux, $\de=1$, \ie
no chemical potential.

Although, to begin with, $q$ is integral, after the sum has been performed to give a
rational function of $q$ one can set $q=1/n$, now with, \eg\ , $n$  integral,
[\pref{Dowcosecs}].

The earliest evaluation known to me is by Jeffery, [\pref{Jeffery}], and I will utilise his
results, as expounded in [\pref{Dowcosecs}]. They are obtained by straightforward
algebra.

The answer is the explicit function of $q$,
  $$
   W_{2g}(q,1)={\pi\over q}{2^{2g}\over\Ga(2g)}\,\sum_{i=0}^{g-1}
   {\Ga(2i+2)\over 2^{2i}}\,A^g_i\,(1-q^{2i+2})\,{\ze(2i+2)\over\pi^{2i+2}}\,,
   \eql{jeff}
  $$
where the $A^g_i$ are constants related to the differentials of nothing, or, equivalently,
to central factorial numbers. They have a combinatorial significance or they can be
evaluated by recursion. They vanish if $i\ge g$ and have been tabulated since 1909. (See
also [\pref{Dowrenexp}].)

To get the energy density, or conformal weight,  according to (\peq{t00}), the combination
$Y(q,1)$ is required,
  $$\eqalign{
  Y(q,1)=&\,W_{2g}(q,1)-{2g-2\over2g-1}\,W_{2g-2}(q,1)\cr
  =&{\pi\,2^{2g}\over\Ga(2g)}\,{1\over q}\sum_{i=0}^{g-1}
   {\Ga(2i+2)\over 2^{2i}}\,B^g_i\,\big(1-q^{2i+2}\big)\,{\ze(2i+2)\over\pi^{2i+2}}\,,
  }
  \eql{whyq}
  $$
using $A_{g-1}^{g-1}=0$ and where the $B^g_i$ are the easily found constants,
  $$
   B^g_i=A^g_i-(g-1)^2A_i^{g-1}\,.
  $$
Note that $B_0^g=0$ which means that there is no term proportional to $q$ in $Y(q,1)$.
This is a consequence of conformal coupling, [\pref{Dowcascone}].

Putting things together and setting $q=1/n$, the conformal weight is,
   $$
   h_n={2^{2g-1}\over(4\pi)^{g-1}}{\Ga(g)\over\Ga(2g)}\,\sum_{i=1}^{g-1}
   {\Ga(2i+2)\over 2^{2i}}\,B^g_i\,\big(n-n^{-2i-1}\big)\,{\ze(2i+2)\over\pi^{2i+2}}\,,
   \eql{hn3}
   $$
which is of the same form as equn.(B.16) in [\pref{Hung}] after a change of summation
variable from $i$ to $j$ where $j=g-1-i$. The constants are different because the methods
are different. The expressions are positive for $n>1$.

In [{\pref{Dowcosecs}], I also computed (following Jeffery) the `fermionic' sum, obtained
by setting $\de=(q-1)/2q$, \ie $\mu=(q-1)/2$, in (\peq{trigsum}). This evaluated to a
sum similar to (\peq{whyq}) except that the Riemann \zf\ was replaced by a Dirichlet
$\eta$--function. I note that a related sum occurs in the spin--half calculations of
[\pref{Hung}].
\section{\bf11. Comments and conclusion}
Explicit expressions for the conformal weights of free scalar twist operators have been
given, and some plotted against chemical potential, for both odd and even dimensions  for
all covering integers. For odd dimensions this last is possible by virtue of an image
relation involving the fluxes along the conical axis.

In even dimensions, polynomials valid for all cone angles can be found. Further, relating
the required energy densities (and, therefore the conformal weights) to a known
trigonometric sum gives a more convenient form to the polynomials, at least for no
chemical potential.

The derivatives of the conformal weights with respect to the chemical potential have
discontinuities indicating phase transitions.

As an alternative to conformally transforming to a hyperbolic $d$--cylinder, the standard
transformation to a conically deformed $d$--sphere can be applied. This has the
advantage of leading to a compact space. The method has been detailed by Belin {\it et
al} [\pref{B}] in the case of S$^3$. Their technique involves finding the eigenvalues and
their degeneracies, then constructing the logdet by direct summation with {\it ad hoc}
zeta function regularisations.

In the language of the orbifolded sphere I have employed before, the sphere is divided
into $2q$ equal lunes of angle $\pi/q$. If $q$ is integral, the lunes properly tile the sphere
but one can extend $q$ into the reals and treat lunes of {\it arbitrary} angle,  even bigger
than $2\pi$. The spectrum on a periodic double lune, of angle $2\pi/q$, is given as the
union of the Dirichlet and Neumann spectra on the single lune. This is best seen by using
the description where the Aharonov--Bohm flux appears in the field equations. The ensuing
analysis will be presented at another time.

The proof of the conjecture of [\pref{BMW}] in section 4 shows the utility, in some
circumstances, of conformally transforming to the flat conical space rather than to the
hyperbolic cylinder.

The considerable simplification effected by Elvang and Hadjiantonis of the integral  to a
sum which has another interpretation would suggest that there is a way of obtaining this
form more directly.

\section{\bf Appendix. A replica equivalent}

For completeness I enlarge on the analysis leading to the decomposition (\peq{images}).
This is only an elaboration of the relation given in [\pref{Dowcascone}] which relies on an
earlier discussion  of quantum field theory on a cone, [\pref{Dowcone}], itself based on a
more general image construction, [\pref{Dowmultc}], on multiply connected manifolds.

I start with an expression, stated for the heat kernel, $K(x,x')$, but valid for any quantity
given linearly in terms of it (such as the Green function, $G$, energy--momentum density,
\etc), on a manifold for which there is a metric containing a conical (two dimensional) part.
Examples are the cosmic string, Rindler, de Sitter, Schwarzschild spaces and spheres, in
the Euclidean case. It is not necessary to know $K$ explicitly to discuss the general
formula. The polar angle of the cone is denoted by $\phi$. The remaining coordinates will
not be displayed and I write the heat--kernel as $K(\phi,\phi')=K(\phi-\phi')$ and set
$\phi'=0$. The metric is assumed SO(2) invariant about the cone axis, which is the SO(2)
fixed--point set. I keep the distance from the cone axis constant so the analysis reduces
to that on the circle.

For a cone, $\phi$ has the range $0\le\phi\le\be$, with end points identified, and then
$K=K_\be$. The polar angle can be unrolled by letting $\be\to\infty$ and one has the
homotopy class relation, [\pref{Dowcone}],
  $$
  K_\be(\phi;\de)=\sum_{m=-\infty}^\infty\,e^{2\pi i m\,\de}\,K_\infty(\phi+m\be;0)\,,
  \eql{hcr}
  $$
for the heat--kernel on a $\be$--cone for a complex field that picks up a phase $\exp(2\pi
i\de)$ on circling the cone once ($\phi\to\phi+\be$), as allowed by general theory on a
multiply connected space. $\de$ can be referred to as a flux, and everything has to be
periodic in it.

Therefore (the limits could be $s=j$ to $s=n-1+j$),
  $$\eqalign{
{1\over n}\sum_{s=0}^{n-1} K_\be(\phi;\de+s/n)&=
\sum_{m=-\infty}^\infty{1\over n}\sum_{s=0}^{n-1}\,e^{2\pi i m\,(\de+s/n)}
\,K_\infty(\phi+m\be;0)\cr
&=\sum_{m=-\infty}^\infty e^{2\pi i m\,\de}
\,K_\infty(\phi+m\be;0)\bigg|_{m=l\,n}\,,\quad l\in\oZ\cr
&=\sum_{l=-\infty}^\infty e^{2\pi i l\,n\,\de}
\,K_\infty(\phi+l\,n\be;0)\bigg|_{m=l\,n}\cr
&=K_{n\be}(\phi;n\de)\,,
}
\eql{proj}
  $$
which is the required decomposition, as given in [\pref{Dowcascone}].\footnote{ From the
corresponding relation for Bernoulli polynomials it could be referred to as a multiplication
or a transformation formula.}

The right--hand side is periodic in $\de$ with period $1/n$, as is the left--hand side, which
follows on shifting the summation limits and bringing them back using the (usual)
periodicity of $K_\be(*;\de)=K_\be(*;\de+1)$. This means that $\de=1/n$ is the same as
$\de=1$ and equivalent to zero chemical potential.

If, for example, $\be=2\pi$, the equation expresses the projection from a (covering) rolled
up circle of circumference $2\pi n$ down to one of circumference $2\pi$ (of the same
radius). For no flux through the big circle, (\peq{proj}) shows that quantities on the
integer covering conical manifold can be expressed as a sum of quantities on a {\it
non}--conical manifold ($\be=2\pi$), but one with a non--zero flux through an axis, as
mentioned in [\pref{Dowcascone}]. Because of the projection, (\peq {proj}) can be
referred to as a (pre--) image sum, \cf\ [\pref{Dowmultc}].

For no flux on the right--hand side  of (\peq{proj}) I need, in my conventions, $n\de=1$,
so that the fluxes on the left--hand side, are $(s+1)/n$ and I can write (\peq{proj}) as,
  $$
  K_{2\pi n}(\phi)={1\over n}\sum_{k=0}^{n-1} K_{2\pi}(\phi\,;k/n)\,,
  \eql{2pin}
  $$
which is, essentially, the result that leads to (\peq{images2}) as used in section 4. Some
subtractions might be necessary to obtain finite quantities.

I emphasise that this decomposition is valid for any manifold that contains an SO(2)
invariant cone.

The same relation holds, at least formally, for the effective action (equivalently free
energy, partition function) and could therefore be regarded as a cheap derivation of the
replica equation,
  $$
   \log Z[n]=\sum _{k=0}^{n-1}\log Z_{k/n}\,,
  $$
\eg\ [\pref{CandH2}] equn.(5), without any diagonalisation of coupled `twist fields'.

The disappearance of the factor $1/n$ is a volume effect as the quantities in
(\peq{images}), and similar relations, refer to densities.

The extension to fermionic fields can be effected by putting $\de\to \de\pm1/2$ in
(\peq{hcr}) to allow for the extra minus sign on circling the $\be$--cone once. This results
in the relation
  $$
  K^f_{n\be}\big(\phi;n(\de+1/2)+1/2\big)={1\over n}
  \sum_{s=0}^{n-1} K^f_\be(\phi;\de+s/n)\,,
  $$
The second argument is the flux through the cone.

Assuming no flux on the left--hand side gives
  $$
 \de\equiv\de_0=-{n-1\over2n}
  $$
so,
  $$
  K^f_{n\be}\big(\phi;1\big)={1\over n}
  \sum_{s=0}^{n-1} K^f_\be(\phi;\de_0+s/n)\,.
  \eql{zcp}
  $$

I refer to this as the zero chemical potential case.

In particular,
  $$
  K^f_{2\pi n}\big(\phi;1\big)={1\over n}
  \sum_{s=0}^{n-1} K^f_{2\pi}(\phi;\de_0+s/n)\,.
  \eql{evenn}
  $$
or, after translating the summation variable from $s$ to $k=s-(n-1)/2$,
  $$\eqalign{
  K^f_{2\pi n}\big(\phi;1\big)={1\over n}\sum_{k=-(n-1)/2}^{(n-1)/2} K^f_{2\pi}(\phi;k/n)
  }
  \eql{oddn}
  $$
Equations (\peq{evenn}) and (\peq{oddn}) replace (\peq{2pin}). If $n=1$ then $s=0$,
$k=0$ and the flux argument of $K_{2\pi}$ is $0$, equivalent to $1$, which is correct by
periodicity.

The partition function equations,
  $$\eqalign{
   \log Z^f[n]&=\sum _{s=0}^{n-1}\log Z^f_{s/n+\de_0}\,\cr
  &\sum _{k=-(n-1)/2}^{(n-1)/2}\log Z^f_{k/n}\,,\cr
  }
  $$
then follow.

For non--zero chemical potential, $\mu$, one sets
  $$
  \de=\mu+\de_0
  $$
and ({\peq{zcp}) is replaced by
  $$\eqalign{
  K^f_{n\be}\big(\phi;n\mu\big)&={1\over n}
  \sum_{s=0}^{n-1} K^f_\be(\phi;\mu+\de_0+s/n)\cr
  &={1\over n}\sum_{k=-(n-1)/2}^{(n-1)/2} K^f_{\be}(\phi;\mu+k/n)
  }
  \eql{zcp}
  $$
which are periodic in $\mu$ with period $1/n$. Hence zero chemical potential can be
achieved by choosing any of $\mu=m/n$, $m\in\oN$.

 \vglue 20truept

 \noin{\bf References.} \vskip5truept
\begin{putreferences}
  \ref{dowtwist}{Dowker,J.S. {\it Conformal weights of charged R\'enyi entropy
  twist operators for free scalars in arbitrary dimensions.} ArXiv:1508.02949.}
   \ref{Dowlensmatvec}{Dowker,J.S. {\it Lens space matter determinants in the vector
   model},  ArXiv:\break 1405.7646.}
   \ref{MandD}{Dowker,J.S. and Mansour,T. {\it J.Geom. and Physics} {\bf 97} (2015) 51.}
   \ref{dowaustin}{Dowker,J.S. 1979 {\it Selected topics in topology and quantum
    field theory}
    \ref{Dowmultc}{Dowker,J.S. \jpa{5}{1972}{936}.}
    (Lectures at Center for Relativity, University of Texas, Austin).}
   \ref{Dowrenexp}{Dowker,J.S. {\it Expansion of R\'enyi entropy for free scalar fields}
   ArXiv:1408.0549.}
   \ref{EandH}{Elvang, H and  Hadjiantonis,M.  {\it Exact results for corner contributions to
   the entanglement entropy and R\'enyi entropies of free bosons and fermions in 3d} ArXiv:
   1506.06729 .}
   \ref{SandS}{T.Souradeep and V.Sahni \prD {46} {1992} {1616}.}
   \ref{CandH}{Casini H., and Huerta,M. \jpa{42}{2009}{504007}.}
   \ref{CandH2}{Casini H., and Huerta,M. J.Stat.Mech {\bf 0512} (2005) 12012.}
   \ref{CandC}{Cardy,J. and Calabrese,P. \jpa{42}{2009}{504005}.}
    \ref{Dow7}{Dowker,J.S. \jpa{25}{1992}{2641}.}
    \ref{Dowcosecs}{Dowker,J.S. {\it On sums of powers of cosecs}, ArXiv:1507.01848.}
    \ref{Jeffery}{Jeffery, H.M. \qjm{6}{1864}{82}.}
   \ref{BMW}{Bueno,P., Myers,R.C. and Witczak--Krempa,W. {\it Universal corner entanglement
   from twist operators} ArXiv:1507.06997.}
  \ref{BMW2}{Bueno,P., Myers,R.C. and Witczak--Krempa,W. {\it Universality of corner
  entanglement in conformal field theories} ArXiv:1505.04804.}
  \ref{BandM}{Bueno,P., Myers,R.C. {\it Universal entanglement for higher dimensional
  cones} ArXiv:1508.00587.}
   \ref{Dowstat}{Dowker,J.S. \jpa{18}{1985}3521.}
   \ref{Dowstring}{Dowker,J.S. {\it Quantum field theory around conical defects} in {\it
   The Formation and Evolution of Cosmic Strings} edited by Gibbons,G.W, Hawking,S.W. and
   Vachaspati,T. (CUP, Cambridge, 1990).}
   \ref{Hung}{Hung,L-Y.,Myers,R.C. and Smolkin,M. {\it JHEP} {\bf 10} (2014) 178.}
\ref{Dow7}{Dowker,J.S. \jpa{25}{1992}{2641}.}
   \ref{B}{Belin,A., Hung,L-Y., Maloney,A., Matsuura,S., Myers,R.C. and Sierens,T.
   {\it JHEP} {\bf 12} (2013) 059.}
   \ref{B2}{Belin,A.,Hung,L-Y.,Maloney,A. and Matsuura,S.
   {\it JHEP01} (2015) 059.}
   \ref{Norlund}{N\"orlund,N.E. \am{43}{1922}{121}.}
    \ref{Norlund1}{N\"orlund,N.E. {\it Differenzenrechnung} (Springer--Verlag, 1924, Berlin.)}
   \ref{Dowconearb}{Dowker,J.S. \pr{36}{1987}{3742}.}
     \ref{Dowren}{Dowker,J.S. \jpamt {46}{2013}{2254}.}
     \ref{DandB}{Dowker,J.S. and Banach,R. \jpa{11}{1978}{2255}.}
     \ref{Dowcen}{Dowker,J.S. {\it Central Differences, Euler numbers and
   symbolic methods} ArXiv: 1305.0500.}
   \ref{Dowcone}{Dowker,J.S. \jpa{10}{1977}{115}.}
   \ref{schulman2}{Schulman,L.S. \jmp{12}{1971}{304}.}
   \ref{DandC}{Dowker,J.S. and Critchley,R. \prD{15}{1977}{1484}.}
     \ref{Thiele}{Thiele,T.N. {\it Interpolationsrechnung} (Teubner, Leipzig, 1909).}
     \ref{Steffensen}{Steffensen,J.F. {\it Interpolation}, (Williams and Wilkins,
    Baltimore, 1927).}
     \ref{Riordan}{Riordan,J. {\it Combinatorial Identities} (Wiley, New York, 1968).}
     \ref{BSSV}{Butzer,P.L., Schmidt,M., Stark,E.L. and Vogt,I. {\it Numer.Funct.Anal.Optim.}
    {\bf 10} (1989) 419.}
      \ref{Dowcascone}{Dowker,J.S. \prD{36}{1987}{3095}.}
      \ref{Stern}{Stern,W. \jram {79}{1875}{67}.}
     \ref{Milgram}{Milgram, M.S., Journ. Maths. (Hindawi) 2013 (2013) 181724.}
     \ref{Perlmutter}{Perlmutter,E. {\it A universal feature of CFT R\'enyi entropy}
     ArXiv:1308.1083 }
     \ref{HMS}{Hung,L.Y., Myers,R.C. and Smolkin,M. {\it Twist operators in
     higher dimensions} ArXiv:1407.6429.}
     \ref{ABD}{Aros,R., Bugini,F. and Diaz,D.E. {\it On the Renyi entropy for
     free conformal fields: holographic and $q$--analog recipes}.ArXiv:1408.1931.}
     \ref{LLPS}{Lee,J., Lewkowicz,A., Perlmutter,E. and Safdi,B.R.{\it R\'enyi entropy.
     stationarity and entanglement of the conformal scalar} ArXiv:1407.7816.}
     \ref{Apps}{Apps,J.S. PhD thesis (University of Manchester, 1996).}
   \ref{CandD}{Candelas,P. and Dowker,J.S. \prD{19}{1979}{2902}.}
    \ref{Hertzberg}{Hertzberg,M.P. \jpa{46}{2013}{015402}.}
     \ref{CaandW}{Callan,C.G. and Wilczek,F. \plb{333}{1994}{55}.}
    \ref{CaandH}{Casini,H. and Huerta,M. \plb{694}{2010}{167}.}
    \ref{Lindelof}{Lindel\"of,E. {\it Le Calcul des Residues} (Gauthier--Villars, Paris,1904).}
    \ref{CaandC}{Calabrese,P. and Cardy,J. {\it J.Stat.Phys.} {\bf 0406} (2004) 002.}
    \ref{MFS}{Metlitski,M.A., Fuertes,C.A. and Sachdev,S. \prB{80}{2009}{115122}.}
    \ref{Gromes}{Gromes, D. \mz{94}{1966}{110}.}
    \ref{Pockels}{Pockels, F. {\it \"Uber die Differentialgleichung $\De
  u+k^2u=0$} (Teubner, Leipzig. 1891).}
   \ref{Diaz}{Diaz,D.E. JHEP {\bf 7} (2008)103.}
  \ref{Minak}{Minakshisundaram,S. {\it J. Ind. Math. Soc.} {\bf 13} (1949) 41.}
    \ref{CaandWe}{Candelas,P. and Weinberg,S. \np{237}{1984}{397}.}
     \ref{Chodos1}{Chodos,A. and Myers,E. \aop{156}{1984}{412}.}
     \ref{ChandD}{Chang,P. and Dowker,J.S. \np{395}{1993}{407}.}
    \ref{LMS}{Lewkowycz,A., Myers,R.C. and Smolkin,M. {\it Observations on
    entanglement entropy in massive QFTs.} ArXiv:1210.6858.}
    \ref{Bierens}{Bierens de Haan,D. {\it Nouvelles tables d'int\'egrales
  d\'efinies}, (P.Engels, Leiden, 1867).}
    \ref{DowGJMS}{Dowker,J.S.  \jpa{44}{2011}{115402}.}
    \ref{Doweven}{Dowker,J.S. {\it Entanglement entropy on even spheres.}
    ArXiv:1009.3854.}
     \ref{Dowodd}{Dowker,J.S. {\it Entanglement entropy on odd spheres.}
     ArXiv:1012.1548.}
    \ref{DeWitt}{DeWitt,B.S. {\it Quantum gravity: the new synthesis} in
    {\it General Relativity} edited by S.W.Hawking and W.Israel (CUP,Cambridge,1979).}
    \ref{Nielsen}{Nielsen,N. {\it Handbuch der Theorie von Gammafunktion}
    (Teubner,Leipzig,1906).}
    \ref{KPSS}{Klebanov,I.R., Pufu,S.S., Sachdev,S. and Safdi,B.R.
    {\it JHEP} 1204 (2012) 074.}
    \ref{KPS2}{Klebanov,I.R., Pufu,S.S. and Safdi,B.R. {\it F-Theorem without
    Supersymmetry} 1105.4598.}
    \ref{KNPS}{Klebanov,I.R., Nishioka,T, Pufu,S.S. and Safdi,B.R. {\it Is Renormalized
     Entanglement Entropy Stationary at RG Fixed Points?} 1207.3360.}
    \ref{Stern}{Stern,W. \jram {79}{1875}{67}.}
    \ref{Gregory}{Gregory, D.F. {\it Examples of the processes of the Differential
    and Integral Calculus} 2nd. Edn (Deighton,Cambridge,1847).}
    \ref{MyandS}{Myers,R.C. and Sinha, A. \prD{82}{2010}{046006}.}
   \ref{RyandT}{Ryu,S. and Takayanagi,T. JHEP {\bf 0608}(2006)045.}
    \ref{Dowcmp}{Dowker,J.S. \cmp{162}{1994}{633}.}
     \ref{Dowjmp}{Dowker,J.S. \jmp{35}{1994}{4989}.}
      \ref{Dowhyp}{Dowker,J.S. \jpa{43}{2010}{445402}.}
       \ref{HandW}{Hertzberg,M.P. and Wilczek,F. \prl{106}{2011}{050404}.}
      \ref{dowkerfp}{Dowker,J.S.\prD{50}{1994}{6369}.}
       \ref{Fursaev}{Fursaev,D.V. \plb{334}{1994}{53}.}
\end{putreferences}

\bye